\documentclass[10pt, conference]{IEEEtran}

\usepackage{cite}
\usepackage{amsmath,amssymb,amsfonts}
\usepackage{amsthm}
\usepackage{mathrsfs}
\usepackage{mathtools}
\usepackage{bbm}
\usepackage{graphicx}
\usepackage{xcolor}
\usepackage{booktabs}
\usepackage{environ}
\usepackage[shortlabels]{enumitem} 
\usepackage{subcaption}

\usepackage{hyperref}

\usepackage{cleveref}

\AtBeginDocument{
\setlength{\belowcaptionskip}{0pt}
\setlength\abovedisplayskip{6pt}
\setlength\belowdisplayskip{6pt}
}

\NewEnviron{myequation}{%
    \begin{align}
    \scalebox{0.5}{$\BODY$}
    \end{align}
    }

\usepackage{color}
\definecolor{dodgerblue4}{rgb}{0.19,0.36,0.56}
\newcommand{\myrule}{\rule[\dimexpr0.5ex - 0.4pt]{1em}{1pt}}

\newcommand{\alice}{\mathcal{A}}
\newcommand{\bob}{\mathcal{B}}

\newcommand{\unknx}{\text{x}}
\newcommand{\colla}{\text{c}}

\newcommand{\co}{\mathit{cont}}
\newcommand{\st}{\mathit{stop}}

\DeclareMathOperator*{\argmax}{argmax}

\makeatletter
\let\oldtheequation\theequation
\renewcommand\tagform@[1]{\maketag@@@{\ignorespaces#1\unskip\@@italiccorr}}
\renewcommand\theequation{(\oldtheequation)}

\makeatother

\begin{document}

\title{A Game-Theoretic Analysis of\\ Cross-Chain Atomic Swaps with HTLCs}

\author{
\IEEEauthorblockN{%
Jiahua Xu%
}
\IEEEauthorblockA{%
\it
UCL Centre for Blockchain Technologies
\\
EPFL
}
\and 
\IEEEauthorblockN{Damien Ackerer}
\IEEEauthorblockA{%
\it
Covario
}
\and
\IEEEauthorblockN{Alevtina Dubovitskaya}
\IEEEauthorblockA{
\it
Lucerne University of Applied Sciences and Arts
\\
Swisscom
\\
UCL Centre for Blockchain Technologies
}
}

\maketitle

\begin{abstract}
To achieve interoperability between unconnected ledgers, hash time lock contracts (HTLCs) are commonly used for cross-chain asset exchange. The solution tolerates transaction failure, and can ``make the best out of worst'' by allowing transacting agents to at least keep their original assets in case of an abort. Nonetheless, as an undesired outcome, reoccurring transaction failures prompt a critical and analytical examination of the protocol. In this study,
we propose a game-theoretic framework to study the strategic behaviors of agents taking part in cross-chain atomic swaps implemented with HTLCs. We study the success rate of the transaction as a function of the exchange rate of the swap, the token price and its volatility, among other variables.
We demonstrate that in an attempt to maximize one's own utility as asset price changes, either agent might withdraw from the swap.
An extension of our model confirms that collateral deposits can improve the transaction success rate, motivating further research towards collateralization without a trusted third party.
A second model variation suggests that a swap is more likely to succeed when agents dynamically adjust the exchange rate in response to price fluctuations.
\end{abstract}


\section{Introduction}
\subsection{Background}
\label{sec:background}
An atomic swap is a coordination task where two parties are willing to exchange assets such that either both parties receive each other's original assets upon successful execution, or nothing in the event of failure~\cite{garcia1983using}.
Atomic swaps are easily achievable on a single ledger by implementing smart contracts such as automated market-making protocols~\cite{xu2021dexAmm}.

For cross-chain asset swaps, the conventional approach is to use a centralized exchange, characterized by high efficiency and transaction speed. 
However, this requires intermediary fees and trust in the exchange (in terms of privacy and transparency of its matching mechanisms). In addition, centralized exchanges are vulnerable to different kinds of attacks \cite{moore2013beware}, from wallet hacking \cite{bitstampHack}, to DDoS attacks \cite{bitfinexDDoS}. 
Over-the-counter (OTC) operations \cite{otcItBit,otcHiveEx}  remain frequent for financial transactions. 
In the financial industry, it is common to use trusted third parties for OTC settlements, such as central clearing counterparties or broker-dealers, that have similar disadvantages as centralized exchanges. 


To address some of the issues with centralized exchanges, and transactions with an intermediary in general, distributed exchanges (DEXs) have recently become a popular tool for cross-chain asset exchange \cite{bisq, decreedGit, komodo, 0x}. 
In such a peer-to-peer (P2P) environment, the transacting agents typically do not know each other and are thus exposed to malicious behaviors from their counterparty in DEXs that only provide match-making services.
Therefore, the major challenge in such settings is to achieve atomicity of the cross-ledger transaction; that is, either the entirety of the transaction is executed, or, in case of failure, nothing occurs \cite{garcia1983using}.

HTLCs,\footnote{The full name may be found with the suffix ``ed'' after hash or lock in research papers. There is no official convention as far as we know.} first proposed on a Bitcoin forum by TierNolan \cite{bitcoinWiki}, have been adopted by some DEXs\cite{decreedGit, komodo} to achieve atomicity of cross-chain transactions without direct communications between the ledgers. Studying and improving HTLCs has since been of high interest \cite{kirsten2018anonymous, zyskind2018enigma, liu2018atomic}. 

A hash time lock contract (HTLC) requires the two agents separately locking their assets on the respective blockchains, using the hash of a secret, generated by one of the users. 
The assets can then be unlocked upon revealing the preimage of a hash. 

\begin{figure}[t]
\centering
\includegraphics[width=0.65\linewidth]{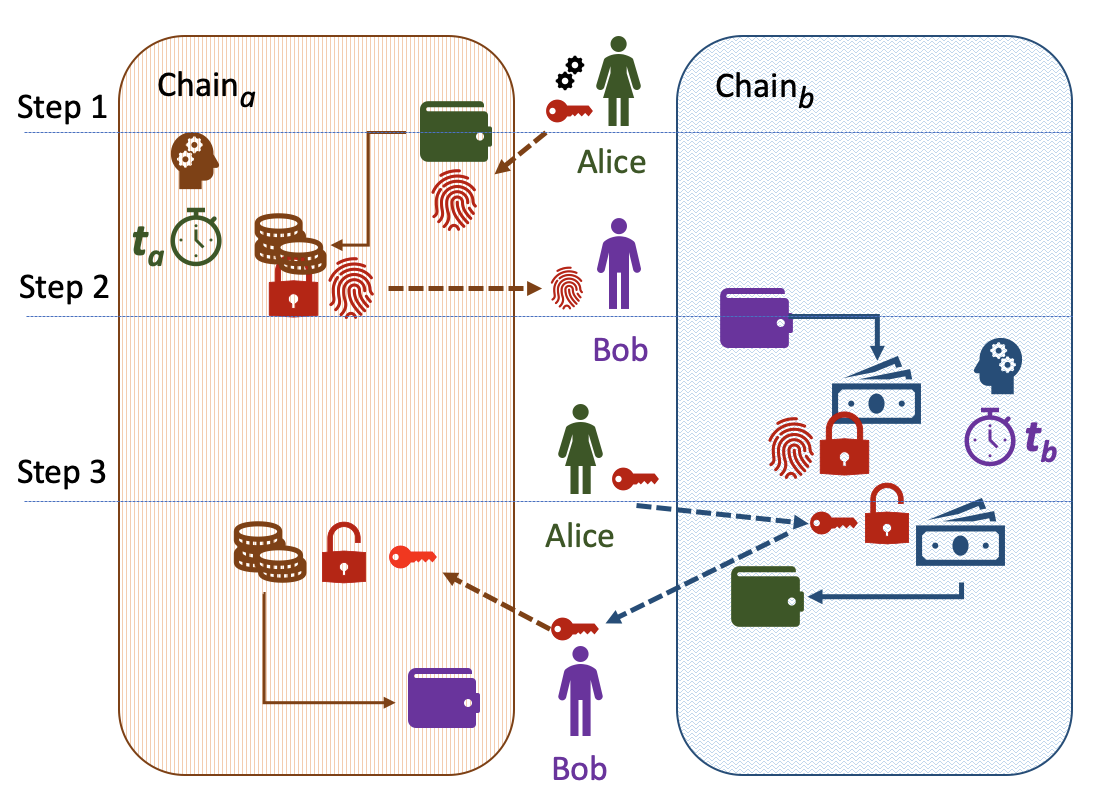}
\caption{Hash time lock contract (HTLC)\label{fig:htlc-f}.}
\end{figure}

The users have accounts (wallets) on two disconnected ledgers (Chain$_a$ and Chain$_b$) executing smart contracts. 
Consider that Alice wants to send assets on Chain$_a$ to Bob, in exchange for assets on Chain$_b$ from Bob. 
At Step 1, Alice initiates the transaction by generating a secret (a key) that will be used to unlock the asset transfers later on. 
She then deploys a smart contract on Chain$_a$, that will lock her assets until time $t_a$. This contract will transfer to Bob the assets only if the secret generated by Alice is revealed and entered into the smart contract.
To verify the secret, Alice reveals its hash as part of the smart contract (cf. \autoref{fig:htlc-f} Step 1). 
One important feature of this contract is that after time $t_a$, should the secret have not been revealed, the smart contract expires and Alice's assets will be unlocked and returned to her wallet.

Next, Bob can verify the contract deployed by Alice on Chain$_a$ (assets, delivery address, etc.) and use the hash submitted by Alice in order to deploy a similar contract on Chain$_b$ (cf. Figure \ref{fig:htlc-f} Step 2), specifying the amount he is willing to transfer to Alice and expiry time $t_b$. Until then Bob's assets are locked on Chain$_b$. 

At Step 3, Alice can verify the contract deployed on Chain$_b$, unlock the assets, and initiate their transfer to her wallet by revealing the secret on Chain$_b$. 
As early as when the secret is revealed in the mempool of Chain$_b$ (even before Alice's transfer is confirmed), Bob can use the secret to unlock the assets on Chain$_a$ and complete the cross-ledger transaction. 
In the best-case scenario, this mechanism enables atomic cross-ledger exchange of the assets without relying on a trusted party and without connection between ledgers. 
If Alice does not unlock the assets on Chain$_b$ before $t_b$, then the assets are transferred back to Bob, thus, she has no incentive to reveal the secret as that would allow Bob to execute the smart contract on Chain$_a$ and transfer the assets to his wallet while keeping his assets on Chain$_b$. 
In turn, once the secret is revealed, Bob's assets are transferred to Alice and he should execute the smart contract on Chain$_a$ immediately in order to complete the transaction, otherwise he transferred his assets without receiving Alice's assets.

\subsection{Contributions}
In this paper, we focus on the standard implementation of a cross-ledger atomic swap with hash time lock contracts, HTLCs between two agents who wish to exchange tokens. In our framework, we assume a stochastic token price and that the counterparties can choose to either {\it continue} or {\it stop} at any stage of the transaction.

We define a game-theoretic framework to study the agents' behaviors and the transaction outcome in atomic swaps.
We focus on the standard protocol with HTLCs, yet the approach can be applied to different setups.
The agents' utility functions depend on 
\begin{enumerate}[1.]
    \item the transaction outcome (success or failure),
    \item the asset price variation (trading profits),
    \item the duration of the transaction (locked in the game).
\end{enumerate}
The agents' utility functions are symmetric, but the agents may have a different idiosyncratic willingness to complete the transaction, the so-called {\it success premium}.
By backward induction, we derive the agents' optimal decisions,
as well as the transaction success rate as a function of the agreed swap rate, actual token price and its volatility, among other variables.
The standard setup has complete information symmetry, and we study the game with uncertainty in counterparties' success premium.

In an extension, we show that if agents would be required to post collateral, everything else being equal, the transaction success rate would be higher.
In another extension, we show that if agents can adjust the amount of tokens that they lock in the HTLCs, everything else being equal, the transaction success rate would be higher.

To the best of our knowledge, our work is the first to perform a thorough step-by-step examination of HTLC agents' behavior through a game-theoretic model with numerical simulations.

\section{Related work} 
\label{subsec:HTLCextensions}

Recent years have witnessed a plethora of cross-ledger transaction solutions besides HTLC.
Wanchain~\cite{wanchain} enables interfacing and asset conversion to the native Wanchain token (Wancoins) in order to perform a cross-ledger asset exchange subsequently. 
Wanchain implements a privacy protection mechanism through a ring signature scheme~\cite{rivest2006leak} and a one-time account mechanism via one-time use wallets created for each transaction. 
Interledger~\cite{thomas2015protocol} uses Byzantine notaries to construct a payment chain from sender to recipient over multiple ledgers, and the STREAM Interledger Transport protocol~\cite{Interledger2020a} apply packetized payments \cite{Dubovitskaya2021}.
Relays~\cite{btcRelay}, sidechains~\cite{back2014enabling, johnson2019sidechains}, off-chain payment channels~\cite{poon2016bitcoin, luu2016secure, miller2017sprites}, and solutions based on chain relays \cite{Zamyatin} require building interfaces to such systems, similar to the case of a blockchain-based medium like Wanchain.

Cross-ledger transaction protocols are actively studied by the distributed ledger community: Borkowski et al. \cite{borkowski2018towards} surveyed atomic swaps for distributed ledgers, Herlihy provided a first extensive analysis of the scheme and demonstrated that HTLCs are still vulnerable to attacks, such as DDoS or secret hack \cite{herlihy2018atomic}. 
Moreover, asset price volatility and malicious behaviour from agents driven by an attempt to maximize financial profits could negatively affect the transaction counterpart. 
For instance, if Alice becomes inactive before completion of the transaction, then the assets will be blocked on both ledgers \cite{Han2019, Zamyatin}. 
While it can be tolerated by an agent, this can incur significant losses to a counterparty.

To reduce the risk of agents being exposed to adverse behaviour, collateral deposits or transaction fees can be used. In a recent work, Han et al.~\cite{Han2019} view atomic swaps as American options (without premium), and discuss ``optionality'' as a risk imposed by the swap initiator. The initiator can choose at any moment before revealing the secret whether to proceed with the swap or to abort it. To reduce the risk of malicious behaviour by the swap initiator, the authors propose to implement a premium mechanism. In our work, we do not define honest or malicious actors explicitly. Instead, we assume that both actors act rationally in their attempts to maximize their utility, and may appear as either ``honest'' or ``malicious'' depending on the movement of the token price. 
One of the trading protocols that support atomic swaps between parties and exchanges is Arwen. Arwen leverages off-chain RFQ trades and uses an escrow-fee mechanism based on blockchain to incentivize a swap initiator to unlock the coins in a timely manner and address lockup griefing in HTLC \cite{heilmanarwen}.

Zamyatin et al. \cite{Zamyatin} suggested posting collateral at least equal to the assets locked on the blockchain for a trade. 
The authors also proposed overcollateralization and a liquidation mechanism to mitigate extreme price fluctuations for both short and long term cross-ledger transactions. 
While such approach reduces economically rational agents' incentive to misbehave, it is disadvantageous in that if an agent would like to transfer all his assets of one kind, he will be obliged to execute multiple transactions, each with an amount (approximately) equal to half the amount of the assets he currently possesses.
In the extension of our model (\Cref{sec:extension}), we also propose that both agents place collateral on one of the chains. We then study the impact of the collateralization on the transaction success rate. This analysis allows us to determine the optimal level of collateral for both agents.

Zakhary et al. \cite{zakhary2019atomic} highlight that even if both participants are honest, the atomicity of HTLC can be violated due to crash failures, preventing smart contract execution before the expiry time of the contract. To address this problem, the authors present all-or-nothing atomic cross-chain commitment protocols and discuss their implementations: the $AC^3TW$-atomic cross-chain commitment protocol with centralized trusted witness and the $AC^3WN$-atomic cross-chain commitment protocol that uses another blockchain as a witness network. This approach shares some similarities with the recently proposed notion of a so-called cross-chain deal \cite{herlihy2019cross}, which, as a generalization of the atomic swap, aims to enhance its expressive power to
support various types of commercial practices. Both cross-chain deals and the protocols proposed in \cite{zakhary2019atomic} are based on exchange of ``proofs'' or ``votes'' instead of relying on hashed time locks. 

Belotti et al. \cite{belotti2019game} are among the first to conduct a game-theoretical analysis of cross-chain swaps (including HTLC as presented in \cite{bitcointalk,herlihy2018atomic} and commitment-based protocols $AC^3TW$ and $AC^3WN$ from \cite{zakhary2019atomic}) and characterise their equilibria. In our work, we focus on the intuition behind the strategic behavior of the participants of HTLC and show that it heavily depends on parameters such as the token price trend and volatility, as well as transaction confirmation time on the employed blockchains.

\section{A game-theoretic analysis}\label{sec:game}


We seek to establish a model with a good balance of simplicity and fidelity. To this end, we apply reasonable and justifiable assumptions, and set realistic parameter values for numeric demonstration.

\subsection{Basic setup} \label{sec:framework}

Alice, denoted by $\alice$, wishes to trade some amount of Token$_a$ for 1 unit of Token$_b$, while Bob, denoted by $\bob$, is willing to do the opposite. Token$_a$ and Token$_b$ are assets from two different blockchains, namely Chain$_a$ and Chain$_b$. $\alice$ and $\bob$ intend to swap assets with each other and agree on the exchange rate:
$P_* \text{ Token}_a = 1 \text{ Token}_b$.
\autoref{tab:balancechange} summarizes the expected balance change of $\alice$'s and $\bob$'s assets on the two chains through the swap.

\begin{table}
\caption{Agents' expected balance change by swap.}
\label{tab:balancechange}
\centering
\begin{tabular}{rrr}
\toprule
            & \multicolumn{2}{c}{\bf Expected balance change by swap}  \\
Agent       & on Chain$_a$      & on Chain$_b$  \\
\cmidrule(lr){1-1} \cmidrule(lr){2-3}
Alice ($\alice$)    & $-P_*$ Token$_a$  & $+1$ Token$_b$\\
Bob ($\bob$)      & $+P_*$ Token$_a$  & $-1$ Token$_b$\\
\bottomrule
\end{tabular}
\end{table}

For simplification, we make the following assumptions:

\begin{enumerate}[1.]
    \item The time it takes for a transaction to be confirmed on Chain$_a$ or Chain$_b$ is constant, equal to $\tau_a$ and $\tau_b$ respectively.
    \item Transaction fees are negligible relative to transaction volume. 

    \item Token$_a$ is the num\'{e}raire, in which Token$_b$ is priced and in which both transacting agents' utilities are measured.
    \label{ass:numeraire}

    \item Token$_b$'s price (denominated in Token$_a$), $P_t$, follows a geometric Brownian motion:
    \label{asm:price}
    \begin{align}
    \label{eq:wiener}
    \ln \tfrac{P_{t + \tau}}{P_{t}} = \left(\mu - \tfrac{\sigma^2}{2}\right) \tau + \sigma \left(W_{t+\tau} - W_t\right)
    \end{align}
    where $W$ follows a Wiener process with drift $\mu$ and infinitesimal variance $\sigma^2$.

    \item $\alice$ and $\bob$ are fully rational, which means they always choose the option that maximizes their utility.

    \item $\alice$ and $\bob$ have the same parametric utility function:
    \label{ass:utilityfunc}
    \begin{align}
    \label{eq:utility}
U_t^i &= \mathbb{E}\left[
\tfrac{(1 + \alpha^i \, S) \,V_{t+T_t^i}}{
e^{r^i \, T_t^i}
}\right]
\end{align}
where

$i$: agent indicator, $i \in \{\alice, \bob\}$

$t$: time when the utility is assessed

$V$: asset value denominated in Token$_a$

$T$: time until end of game, when no further events directly connected to swap will occur

$r$: discount rate, $r>0$

$S$: success indicator, 1 if the swap succeeds (i.e. agents balance change follows \autoref{tab:balancechange}), 0 if fails

$\alpha$: success premium

\item $\alice$ and $\bob$ are aware of the value of each other's parameter set, i.e. $\alice$ knows $(r^\bob, \alpha^\bob)$, and $\bob$ knows $(r^\alice, \alpha^\alice)$.
\label{ass:agentpara}
\end{enumerate}

According to \autoref{eq:wiener}, given Token$_b$'s price at time~$t$, $P_t$, the expectation (denoted by $\mathcal{E}$), probability density function (PDF, denoted by $\mathcal{P}$), and cumulative density function (CDF, denoted by $\mathcal{C}$) of its price at $t+\tau$ can be expressed as:
\begin{align}
\mathcal{E}(P_{t},\tau) & \coloneqq
\mathbb{E}[P_{t+ \tau} \,|\, P_t]
=
P_{t} e^{\mu \tau}
\nonumber \\
\mathcal{P}(x,P_{t},\tau)
& \coloneqq
\mathbb{P}[P_{t+ \tau} = x \,|\, P_{t}]
=
\tfrac{e^{-\frac{\left(
\ln \frac{x}{P_t} -
\left(\mu - \frac{\sigma^2}{2} \right) \tau
\right)^2}{2\tau \sigma^2}}}{\sqrt{2\pi \tau} \sigma x}
\nonumber \\
\mathcal{C}(x,P_{t},\tau)
& \coloneqq
\mathbb{P}[P_{t+ \tau} \leq x \,|\, P_{t}]
=
\tfrac{\operatorname{erfc}\big(\frac{
\ln \frac{x}{P_t} -
\left(\mu - \frac{\sigma^2}{2} \right) \tau
}{\sqrt{2\tau} \sigma}
\big)}{2}
\nonumber 
\end{align} where $\operatorname{erfc}$ is the complementary error function, and $x>0$.

All actors act rationally to maximize their utility as defined in \autoref{eq:utility}. Intuitively, actors with a higher success premium $\alpha$ will act more ``honestly'', i.e. ceteris paribus, they are more likely to continue the game; on the other hand, actors with a lower success premium $\alpha$ may appear ``malicious'', since ceteris paribus, they are more likely to withdraw from the game.

\begin{table}
\centering
\caption{Notations summary. \label{tab:notation}}
\setlength{\tabcolsep}{2pt}
\footnotesize
\begin{tabular}{@{}c|l@{}}
\toprule
Notation & Description \\
\midrule
$\alice$, $\bob$ & Alice, Bob\\
$\tau_a$, $\tau_b$ & Transaction confirmation time on Chain$_a$, Chain$_b$\\
$\varepsilon_b$ & Time for an initiated transaction to become discoverable \\ & in the mempool of Chain$_b$  \\
$t$ & Point in time \\
$t_a$, $t_b$ & Points in time when the HTLCs on Chain$_a$, Chain$_b$ expire \\
$P$ & Price of Token$_b$ denominated in Token$_a$\\
$P_*$ & Agreed price of Token$_b$ denominated in Token$_a$ \\
$U$ & Agent's utility denominated in Token$_a$ \\
$V$ & Asset value denominated in Token$_a$ \\
$r$ & Discount rate representing time preference \\
$S$ & Indicator of whether the swap succeeds ($=1$) or not ($=0$) \\
$\alpha$ & Success premium \\
$T$ & Time until end of game \\
$\mu$ & Wiener Process drift, see \autoref{eq:wiener} \\
$\sigma^2$ & Wiener Process variance, see \autoref{eq:wiener} \\
$\mathcal{E}(P_t,\tau)$ & Expectation of Token$_b$ price at $t+\tau$ given its time-$t$ price $P_t$\\
$\mathcal{P}(\cdot, P_t, \tau)$ & PDF of Token$_b$ price at $t+\tau$ given its time-$t$ price $P_t$\\
$\mathcal{C}(\cdot, P_t, \tau)$ & CDF of Token$_b$ price at $t+\tau$ given its time-$t$ price $P_t$\\
\bottomrule
\end{tabular}
\end{table}

\subsection{Decision timeline} \label{sec:timeline}

Let $\varepsilon_b$ denote the time needed to look up a transaction in the mempool of Chain$_b$ after it has been initiated. This time is smaller than the transaction confirmation time on Chain$_b$, i.e.
\begin{equation}
\label{eq:epsilon}
\varepsilon_b < \tau_b
\end{equation}

Let $\{t_n\}_{n\in \mathbb{N}}$ denote the points in time when agents have to make a decision, and $t_a$ ($t_b$) denote the point in time when the HTLC on Chain$_a$ (Chain$_b$) expires. See \autoref{tab:notation} for a comprehensive list of notations used in this paper.

According the HTLC protocol described in Section~\ref{sec:background}, an atomic swap should work as follows:

\subsubsection{Agreement and preparation}
    \paragraph*{$t_0$}
     $\alice$ and $\bob$ agree on the swap conditions, including exchange rate $P_*$, contract lock expiration time $t_a$ and $t_b$ etc. $\alice$ generates a secret and its hash.

\subsubsection{Action}
\label{sec:action}
    \paragraph*{$t_1$}
    $\alice$ uses the hash generated at $t_0$ to lock $P_*$ Token$_a$ on Chain$_a$ through an HTLC that expires at $t_a$; thus,
    \begin{equation}
    t_1 \geq t_0
    \label{eq:t1}
    \end{equation}

    \paragraph*{$t_2$}
    $\bob$ uses the same hash to lock 1 Token$_b$ on Chain$_b$ through another HTLC that expires at $t_b$.

    $\bob$ does so only after verifying that $\alice$'s contract is in order and that its deployment has been confirmed on Chain$_a$; thus,
    \begin{equation}
    t_2 \geq t_1 + \tau_a
    \label{eq:t2}
    \end{equation}

    \paragraph*{$t_3$}
    $\alice$ uses the secret to unlock the 1 Token$_b$ on Chain$_b$.
    $\alice$ does so only after verifying that $\bob$'s contract is in order and that its deployment has been confirmed on Chain$_b$; thus,
    \begin{equation}
    t_3 \geq t_2 + \tau_b
    \end{equation}

    \paragraph*{At $t_4$}
    $\bob$ uses the same secret to unlock the $P_*$ Token$_a$ on Chain$_a$.
    $\bob$ does so only after seeing the secret been revealed by $\alice$ in the mempool of Chain$_b$; thus
    \begin{equation}
    t_4 \geq t_3 + \tau_\varepsilon
    \end{equation}

\subsubsection{Receipt}~%

\noindent
Which token an agent receives, and when, depends on the outcome of the swap. If both $\alice$ and $\bob$ hold on to their agreement by following the steps during the action phase as described in Section~\ref{sec:action}, then the swap succeeds and $\alice$ and $\bob$ receive tokens at the following two points in time respectively:
    \paragraph*{$t_5$}
    $\alice$ receives the 1 Token$_b$ after her transaction is confirmed on Chain$_b$, and this must take place before the lock contract expires at $t_b$; thus
    \begin{equation}
    t_5 = t_3 + \tau_b \leq t_b
    \end{equation}

    \paragraph*{$t_6$}
     $\bob$ receives the $P_*$ Token$_a$ after his transaction is confirmed on Chain$_a$, and this must take place before the lock contract expires at $t_a$; thus
     \begin{equation}
     t_6 = t_4 + \tau_a \leq t_a
     \label{eq:t6}
     \end{equation}


If, however, $\alice$ or $\bob$ withdraws at any point during the action phase as described in \ref{sec:action}, then the swap fails and $\alice$ and $\bob$ have their original tokens returned to them at the following two points in time respectively:

    \paragraph*{$t_7$}
    the HTLC on Chain$_b$ returns $\bob$'s original 1 Token$_b$ to him when the time lock expires at $t_b$, and $\bob$ receives the 1 Token$_b$ at $t_7$; thus
    \begin{equation}
    t_7 = t_b + \tau_b
    \end{equation}

    \paragraph*{$t_8$}
    the HTLC on Chain$_a$ returns $\alice$'s original 1 Token$_a$ to her when the time lock expires at $t_a$, and $\alice$ receives the $P_*$ Token$_a$ at $t_8$; thus
    \begin{equation}
    t_8 = t_a + \tau_a
     \label{eq:t8}
     \end{equation}


\begin{figure}[tb]
\begin{subfigure}{\linewidth}
\centering
\includegraphics[
width=0.8\linewidth,
trim = {0, 37, 0, 0}]
{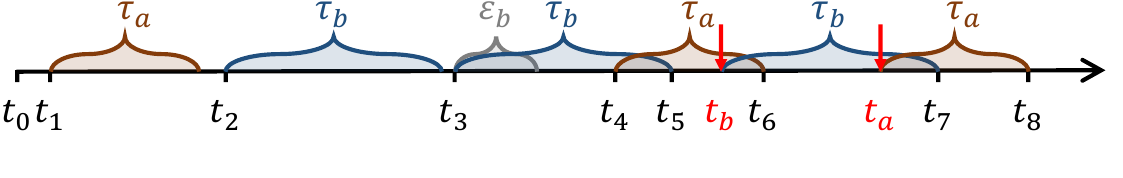}
\caption{Timeline with arbitrary amount of waiting time\label{fig:tl1}}
\end{subfigure}
\par\bigskip
\begin{subfigure}{\linewidth}
\centering
\includegraphics[width=0.8\linewidth,
trim = {0, 20, 0, 0}]{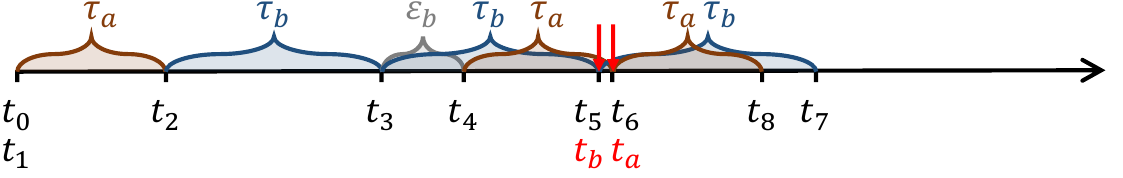}
\caption{Idealized timeline with zero waiting time\label{fig:tl2}}
\end{subfigure}
\caption{Swap timeline as described in Section \ref{sec:timeline}.}
\end{figure}

Combining \ref{eq:epsilon}--\ref{eq:t8}, we get
\begin{equation}
    \begin{cases}
    t_0 \leq t_1 < t_1 + \tau_a \leq t_2 < t_2+\tau_b \leq t_3 < t_3 + \varepsilon_b
    \\
    t_3 + \varepsilon_b < t_3 + \tau_b = t_5 \leq t_b < t_b + \tau_b =t_7
    \\
    t_3 + \varepsilon_b \leq t_4 < t_4+\tau_a = t_6 \leq t_a < t_a + \tau_a = t_8
    \end{cases}
\end{equation}

The relationship between different points in time can thus be illustrated as \autoref{fig:tl1}.

\subsection{Zero waiting time}

An idealized decision-making time of 0 allows the game to be characterized as a discrete one, where actions can only be taken at a specified, finite set of points in time. In this way, we can express the relationships between the critical points in time as (\autoref{fig:tl2}):
\begin{align}
    \begin{cases}
    t_5 = \underbrace{
    \overbrace{t_1}^{t_0} + \tau_a
    }_{t_2} + \tau_b + \tau_b = t_b & t_7 = t_b + \tau_b
    \\
    t_6 = \underbrace{
    \overbrace{t_2 + \tau_b
    }^{
    t_3
    }+ \varepsilon_b
    }_{t_4} + \tau_a = t_a & t_8 = t_a + \tau_a
    \end{cases}
    \label{eq:ideatime}
\end{align}

This model simplification can be justified for multiple reasons. 
Firstly, at the outset of the swap, it should be of both agents' interest to agree on the terms such that the swap can be carried out in a swift manner. From a game-theoretical perspective, lengthening the decision-making time increases optionality for $\alice$ at $t_3$, thus reducing the expected utility for $\bob$. $\bob$ in turn would postpone his decision at $t_2$ until as late as possible, to maximize his own optionality and minimize $\alice$'s future optionality. $\alice$ in turn would wait as long as possible at $t_1$ to kick off the swap in order to minimize $\bob$'s future optionality and maximize her own. In turn at $t_0$, $\bob$ is incentivised to only agree to the shortest time possible to reduce $\alice$'s optionality at $t_1$.

In addition, a long lock time can reduce liquidity for $\alice$ and $\bob$ collectively. Formally, the negative impact of lock time on agents' utility is captured by their positive discount rate $r$, as shown in \ref{eq:utility}.  Therefore, the agents would set the contract expiration time as early as possible, to reduce the time of assets being locked, in two ways:
\begin{enumerate}[1.]
    \item An agent can receive his/her counterparty's original asset earlier rather than later in case the swap eventually succeeds. This is because an agent is forced to choose whether to continue or to withdraw immediately each time it is his/her turn to take an action; no immediate action (i.e. waiting) is equivalent to withdrawal since it precludes timely completion of necessary transactions before the contract expiration time, and thus lead to failure of the swap.
    \item An agent can get back his/her original asset as soon as the counterparty withdraws from the swap, i.e. when the swap eventually fails.
\end{enumerate}

\subsection{Default parameters} \label{sec:defaultvalue}

We use backward induction to derive the optimal strategy for each agent. We solve for agents' best strategy numerically and graphically, as we show later in \Cref{sec:backwardation} that it quickly becomes non-trivial to analytically derive a closed-form expression. To this end, we set the default value of certain parameters as in \autoref{tab:default}. We additionally specify the unit of parameters to put the model into perspective. In the following, we discuss the plausibility of selected values for blockchain-specific parameters.

\paragraph*{Transaction confirmation time $\tau$}  We set the confirmation time on both chains to be in the order of hours. Confirmation time in this paper refers to the time needed to reach transaction finality with a high probability, which typically equals a multiple of the block time.
Given the wide adoption of the computationally heavy consensus mechanism---Proof of Work \cite{Ferdous2020}, it is to date still common for a blockchain to have an hour-long confirmation time.\footnote{See e.g. \url{https://support.kraken.com/hc/en-us/articles/203325283-Cryptocurrency-deposit-processing-times}}

\paragraph*{Price trend $\mu$} The default value of a positive price trend suggests the deflationary nature of Token$_b$, for example caused by higher levels of token buyback and burn compared to Token$_a$ \cite{Tang2018}. In \Cref{sec:sucrate}, we explore the possibility when  Token$_b$ is inflationary, i.e. $\mu < 0$, and when $\mu = 0$.

\paragraph*{Volatility $\sigma$} The default hourly volatility value of 10\% aligns with empirical evidence \cite{Digiconomist2014}.

In \Cref{sec:sucrate}, we inspect modelling results with an array of different values for each parameter.

\begin{table}[t]
\centering
\tiny
\caption{Default value of parameters. \label{tab:default}}
\hrule
\begin{align*}
\alpha^\alice  & = 0.3 &
r^\alice & = 0.01\ (/\text{hour}) &
\tau_a  & = 3\ (\text{hour})&
\varepsilon_b  & = 1\ (\text{hour}) &
\mu  & = 0.002\ (/\text{hour}) \\
\alpha^\bob  & =  0.3 &
r^\bob  & = 0.01\ (/\text{hour}) &
\tau_b  & = 4\ (\text{hour}) &
P_{t_0} & = 2\ (\text{Token}_a) &
\sigma  & = 0.1\ (/\sqrt{\text{hour}})
\end{align*}
\hrule
\end{table}

As per assumption, the values of all the parameters displayed in \autoref{tab:default} are {\em common knowledge} (i.e. $\alice$ knows, $\bob$ knows, and $\alice$ knows that $\bob$ knows etc).

\subsection{Backward induction} \label{sec:backwardation}

With backward induction, we start from $t_4$, the last possible action point. We then move backward to an earlier action point each time, assuming the swap is still ongoing (i.e. nobody has withdrawn up until that point). Recall that we use the idealized framework where decision-making time is reduced to a point in time. That is, at each decision-making point in time, agents choose an action from the two-element action set $\{\co, \st\}$.

\subsubsection{$t_4$}\label{sec:t4}
$\bob$ decides whether to unlock Token$_a$ ($\co$) or not ($\st$).
Once $\alice$ has unlocked Token$_b$ with her pre-generated secret, it does not make sense for $\bob$ to withdraw and forgo the locked Token$_a$ (which yields to zero utility). Therefore, as soon as $\bob$ sees the secret revealed through $\alice$'s transaction from Chain$_b$'s mempool, $\bob$ uses the secret to unlock Token$_a$. Thus, $\bob$ chooses to continue {\em with certainty}.

\subsubsection{$t_3$}\label{sec:t3}
$\alice$ decides whether to unlock Token$_b$ ($\co$) or not ($\st$).

\paragraph*{$\co$}
$\alice$ unlocks the 1 Token$_b$, and receives it at $t_5 = t_3+\tau_b$, in which case $\bob$ gets $P_*$ Token$_a$ at $t_6=t_3+(\varepsilon_b+\tau_a)$.
\begin{align}
U^\alice_{t_3}(\co) & =
\tfrac{(1+\alpha^{\alice})\mathcal{E}(P_{t_3}, \tau_b)}{e^{r^\alice \tau_b}}
\label{eq:At3cont}
\\
U^\bob_{t_3}(\co) & =
\tfrac{(1+\alpha^{\bob})P_*}{e^{r^\bob (\varepsilon_b+\tau_a)}}
\label{eq:Bt3cont}
\end{align}

\paragraph*{$\st$} 
$\alice$ waives the contract and has the $P_*$ Token$_a$ returned to her at $t_8 = t_3 + (\varepsilon_b + 2 \tau_a)$, in which case $\bob$ gets 1 Token$_b$ at $t_7 = t_3+2\tau_b$.
\begin{align}
U^\alice_{t_3}(\st) & = \tfrac{P_*}{
e^{r^\alice(\varepsilon_b + 2\tau_a})
}
\label{eq:At3stop}
\\
U^\bob_{t_3}(\st) & =
\tfrac{
\mathcal{E}(P_{t_3},2\tau_b)
}{e^{r^\bob \, 2\tau_b}}
\label{eq:Bt3stop}
\end{align}

\begin{figure}
\includegraphics[
    height=0.135\textheight,
    trim = {52, 40, 25, 0}, clip
    ]{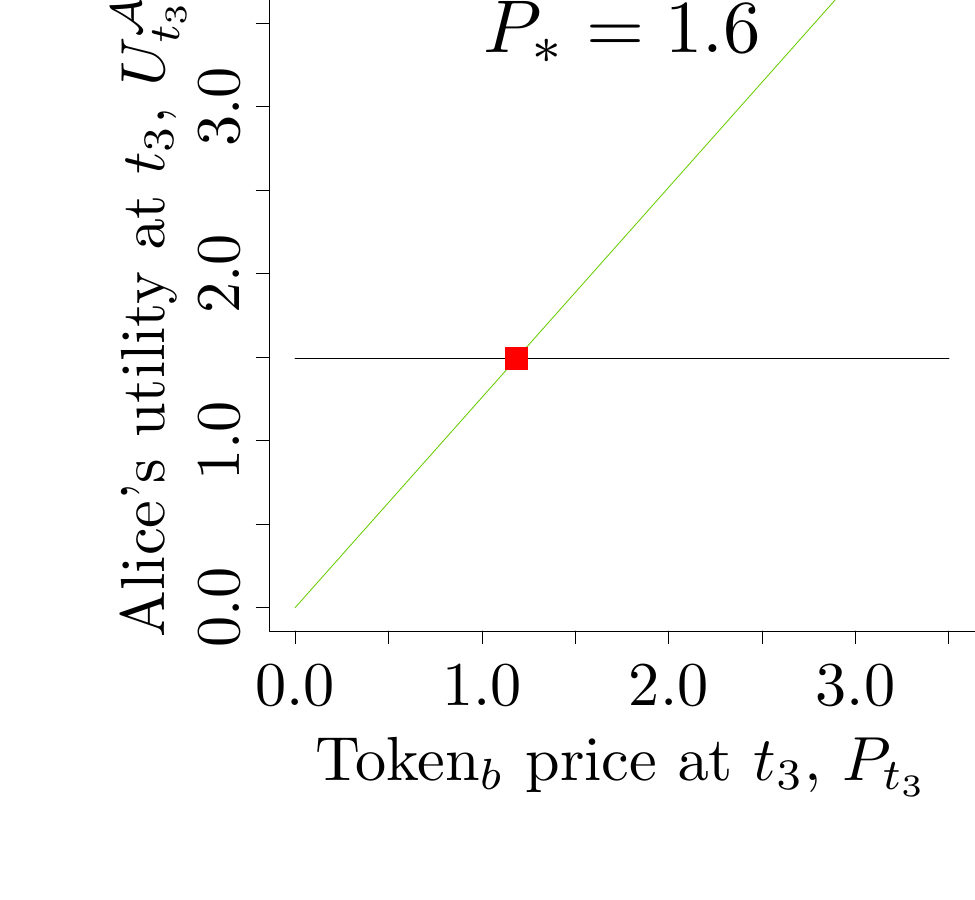}
~\includegraphics[
    height=0.135\textheight,
    trim = {120, 40, 25, 0}, clip
    ]{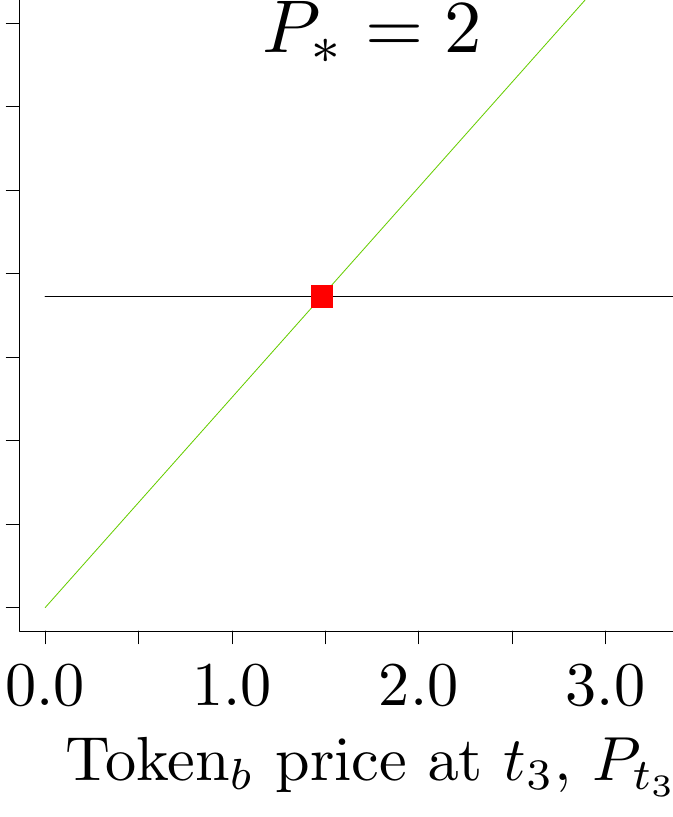}
~\includegraphics[
    height=0.135\textheight,
    trim = {120, 40, 15, 0}, clip
    ]{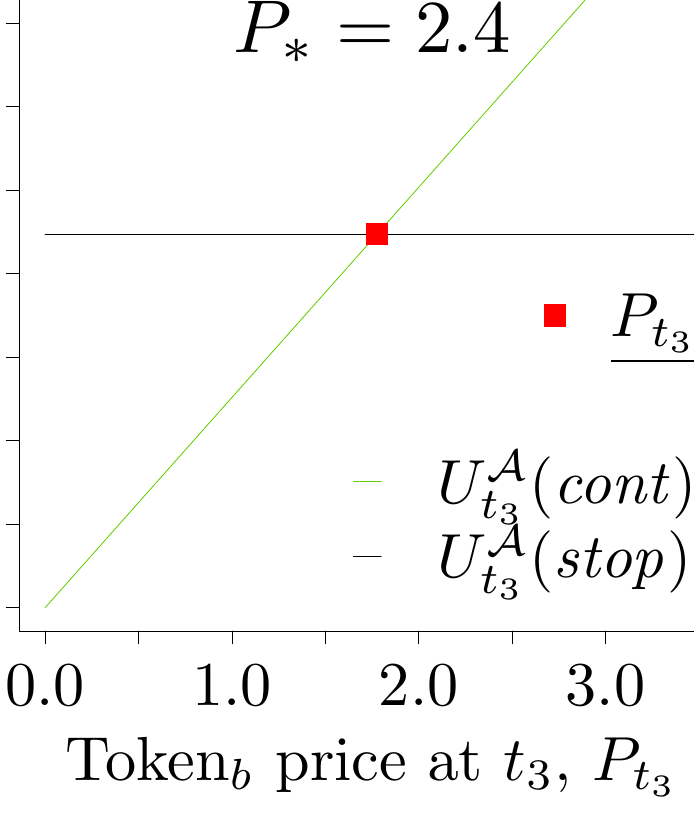}
    \caption{Alice's utility at $t_3$, ($U_{t_3}^\alice$) with different strategies ($\co, \st$), Token$_b$ price values ($P_{t_3}$), and exchange rate values ($P_*$).}
    \label{fig:util3_A}
\end{figure}

By assumption, $\alice$ chooses the option that maximizes her utility. Intuitively,
    when current Token$_b$ price $P_{t_3}$ is sufficiently large, $\alice$ chooses to stop so that she receives Token$_b$ in the end;
    when $P_{t_3}$ is sufficiently small, $\alice$ chooses to continue so that she receives Token$_a$ in the end.

Let $\underline{P_{t_3}}$ denote the cut-off price that equates $U^\alice_{t_3}(\co)$ and $U^\alice_{t_3}(\st)$, i.e.
\begin{align}
\tfrac{(1+\alpha^{\alice})\mathcal{E}(\underline{P_{t_3}}, \tau_b)}{e^{r^\alice \tau_b}}
& =
\tfrac{P_*}{
e^{r^\alice(\varepsilon_b + 2\tau_a})
}
\nonumber \\
\underline{P_{t_3}} &=
\tfrac{e^{
(r^\alice - \mu) \tau_b - r^\alice(\varepsilon_b + 2\tau_a)
} P_*}{
1+\alpha^\alice
}
\label{eq:p3lower}
\end{align}

Clearly,
$\underline{P_{t_3}}$ increases with $P_*$ (see also \autoref{fig:util3_A}). This is because higher $P_*$ makes the option stop more attractive for $\alice$, driving the threshold price $\underline{P_{t_3}}$ higher.

$\alice$'s strategy at $t_3$ can be summarized as:
\begin{equation}
    \begin{cases}
    \co, & P_{t_3} > \underline{P_{t_3}}\\
    \st, & P_{t_3} \leq \underline{P_{t_3}}
    \end{cases}
\end{equation}

\subsubsection{$t_2$}
$\bob$ decides whether to write an HTLC on Chain$_b$ ($\co$) or not ($\st$).
\label{sec:t2}

\paragraph*{$\co$}
Even when $\bob$ chooses to continue, whether the swap eventually succeeds depends on how Token$_b$ price evolves until $t_3$. Therefore, the utility of $\alice$ and $\bob$ at $t_2$ can be expressed with their time-discounted, expected utility at~$t_3$:
\begin{align}
U^\alice_{t_2}(\co)
& =
\tfrac{{
\int_{\underline{P_{t_3}}}^{\infty}
\mathcal{P}\left(x ,P_{t_2},\tau_b\right)
U^\alice_{t_3}(\co) dx 
\atop +
\mathcal{C}(\underline{P_{t_3}} ,P_{t_2},\tau_b)
U^\alice_{t_3}(\st)
}}{
e^{r^{\alice}\tau_b}
}
\label{eq:At2cont}
\\
U^\bob_{t_2}(\co)
& =
\tfrac{{
\left[
1-\mathcal{C}(\underline{P_{t_3}} ,P_{t_2},\tau_b)
\right]
U^\bob_{t_3}(\co)
\atop +
\int_0^{\underline{P_{t_3}}}
\mathcal{P}\left(x ,P_{t_2},\tau_b\right)
U^\bob_{t_3}(\st)dx
}}{
e^{r^{\bob}\tau_b}
}
\label{eq:Bt2cont}
\end{align}

\paragraph*{$\st$}
$\bob$ withdraws from the deal and keeps 1 Token$_b$ at $t_2$. Since $\alice$ already locked $P_*$ Token$_a$, she will receive her original Token$_a$ back at $t_8 = t_2 + (\tau_b + \varepsilon_b + 2\tau_a)$.
\begin{align}
U^\alice_{t_2}(\st) & = \tfrac{P_*}{
e^{r^\alice(\tau_b + \varepsilon_b + 2\tau_a)}
}
\label{eq:At2stop}
\\
U^\bob_{t_2}(\st) & =
P_{t_2}
\label{eq:Bt2stop}
\end{align}

By assumption, $\bob$ also chooses the option that maximizes his utility. In \autoref{fig:util3_A}, $U^\bob_{t_3}(\co)$ and $U^\bob_{t_3}(\st)$ are plotted as a function of $P_{t_3}$. Intuitively,
    if current Token$_b$ price $P_{t_2}$ is very low, then although $\bob$ would want to swap to receive Token$_a$, he can safely assume that Alice would not honor the agreement at $t_3$ anyway, so $\bob$ will not bother to continue;
    if $P_{t_2}$ is very high, then $\bob$ would like to keep Token$_b$ and won't swap at all.

\begin{figure}
\includegraphics[
    height=0.135\textheight,
    trim = {52, 40, 25, 0}, clip
    ]{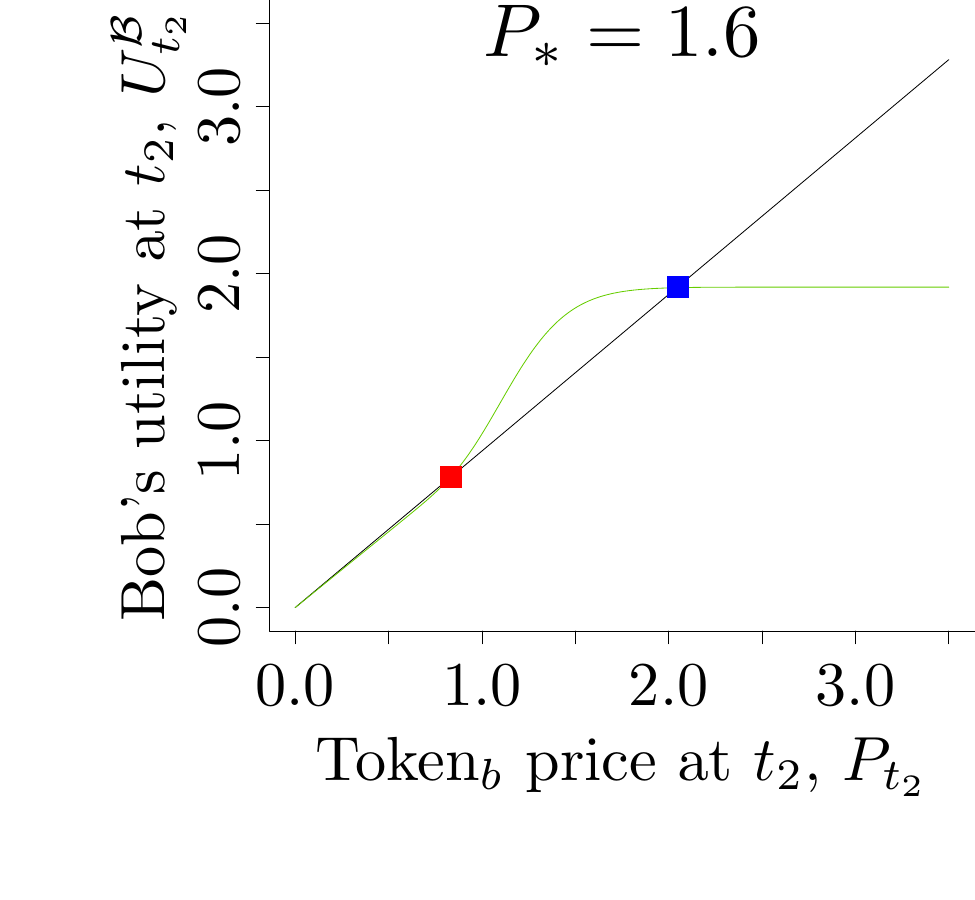}
~\includegraphics[
    height=0.135\textheight,
    trim = {120, 40, 25, 0}, clip
    ]{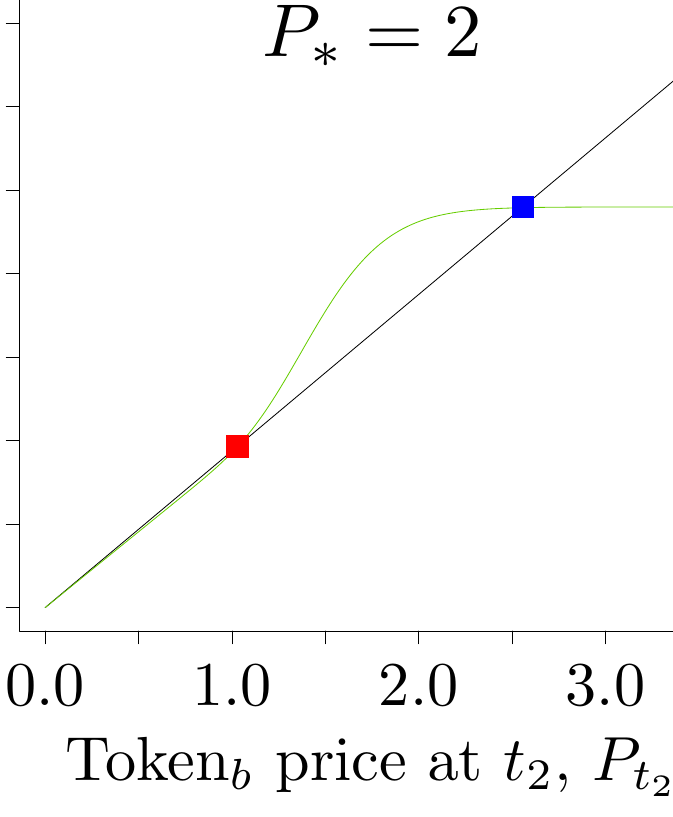}
~\includegraphics[
    height=0.135\textheight,
    trim = {120, 40, 15, 0}, clip
    ]{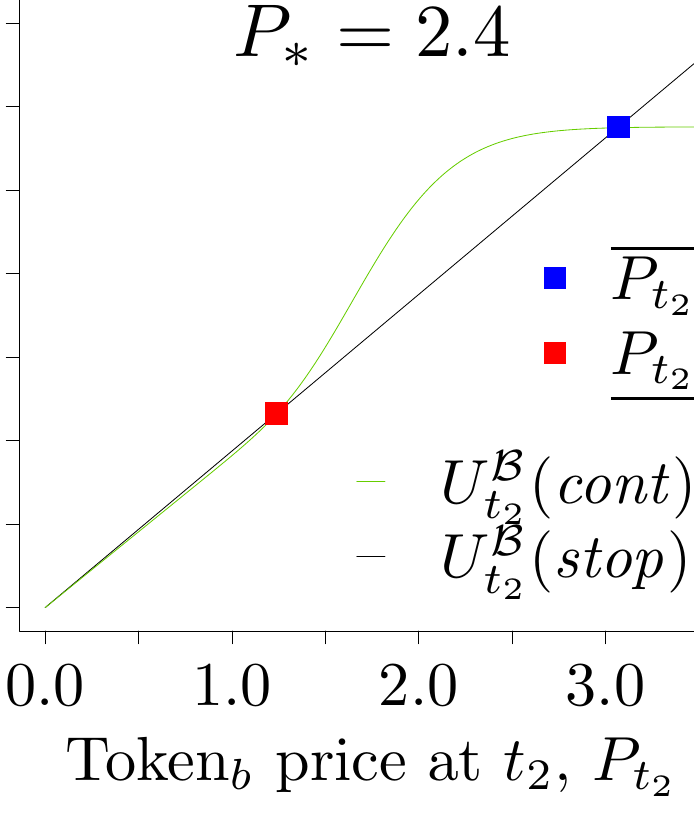}
    \caption{Bob's utility at $t_2$, ($U_{t_2}^\bob$) with different strategies ($\co, \st$), Token$_b$ price values ($P_{t_2}$), and exchange rate values ($P_*$).}
    \label{fig:util3_B}
\end{figure}

Therefore, Bob chooses $\co$ when $P_{t_2}$ falls in a feasible range, denoted by $(\underline{P_{t_2}}, \overline{P_{t_2}})$. \autoref{fig:util3_B} shows that this range expands and shifts to the higher end with larger $P_*$.

$\bob$'s strategy at $t_2$ can be summarized as:
\begin{equation}
    \begin{cases}
    \co, &  \underline{P_{t_2}} < P_{t_2} \leq  \overline{P_{t_2}}\\
    \st, & P_{t_2} \leq \underline{P_{t_2}} \text{ or } P_{t_2} > \overline{P_{t_2}}
    \end{cases}
\end{equation}

Note that depending on the value of parameters, the two utility curves might not always have two intersections (disregarding the origin). For example, the lower $\alpha^\bob$ is, the narrower the feasible range of $P_{t_2}$ is, because $\bob$ is less desperate to swap. When $\alpha^\bob$ is sufficiently small, $U^\bob_{t_2}(\co)<
U^\bob_{t_2}(\st), \forall P_{t_2}>0$, and the swap always fails. This will be further discussed in Section \ref{sec:sucrate}.

\subsubsection{$t_1$}
\label{sec:t1}

$\alice$ decides whether to initiate the swap by writing an HTLC on Chain$_a$ ($\co$) or not ($\st$).

\paragraph*{$\co$} The utility of $\alice$ and $\bob$ at $t_1$ can be expressed by time-discounting their expected utility at ${t_2 = t_1 + \tau_a}$:
\begin{align}
U^\alice_{t_1}(\co)
& =
\tfrac{{
\int_{\underline{P_{t_2}}}^{\overline{P_{t_2}}}
\mathcal{P}(P_{t_2} ,P_{t_1},\tau_a)
U^\alice_{t_2}(\co) dP_{t_2} +
\atop
\Big[ 1-
\mathcal{C}(\overline{P_{t_2}} ,P_{t_1},\tau_a) + 
\mathcal{C}(\underline{P_{t_2}} ,P_{t_1},\tau_a)
\Big]
U^\alice_{t_2}(\st)
}}{
e^{r^{\alice}\tau_a}
}
\label{eq:At1cont} \\
U^\bob_{t_1}(\co) 
& =
\tfrac{{
\int_{\underline{P_{t_2}}}^{\overline{P_{t_2}}}
\mathcal{P}(P_{t_2} ,P_{t_1},\tau_a)
U^\bob_{t_2}(\co)  dP_{t_2} +
\atop
\int_0^{\underline{P_{t_2}}}
\mathcal{P}(P_{t_2} ,P_{t_1},\tau_a)
U^\bob_{t_2}(\st)dP_{t_2}
}}{
e^{r^{\bob}\tau_a}
}
\end{align}

\paragraph*{$\st$}  $\alice$ does not initiate the swap and keeps $P_*$ Token$_a$ at $t_1$. $\bob$ also keeps his 1 Token$_b$.
\begin{align}
U^\alice_{t_1}(\st) & = P_*
\label{eq:At1stop}\\
U^\bob_{t_1}(\st) & =
P_{t_1}
\end{align}

Recall that in our idealized swap game, $\alice$ must initiate the swap {\em immediately} after the terms of swap are agreed upon, as allowing for waiting time only reduces agents' utility. Therefore, $\alice$ and $\bob$ must agree on a rate $P_*$ that makes $\alice$ willing to take the first step. Intuitively,
    if $P_*$ is too high, then $\alice$ would not want to swap;
    if $P_*$ is too low, then $\alice$ understands the high likelihood of fail because $\bob$ would not want to continue at $t_2$, so $\alice$ would not start the swap either.

\begin{figure}
\centering
\includegraphics[
    height=0.135\textheight,
    trim = {52, 40, 15, 0}, clip
    ]{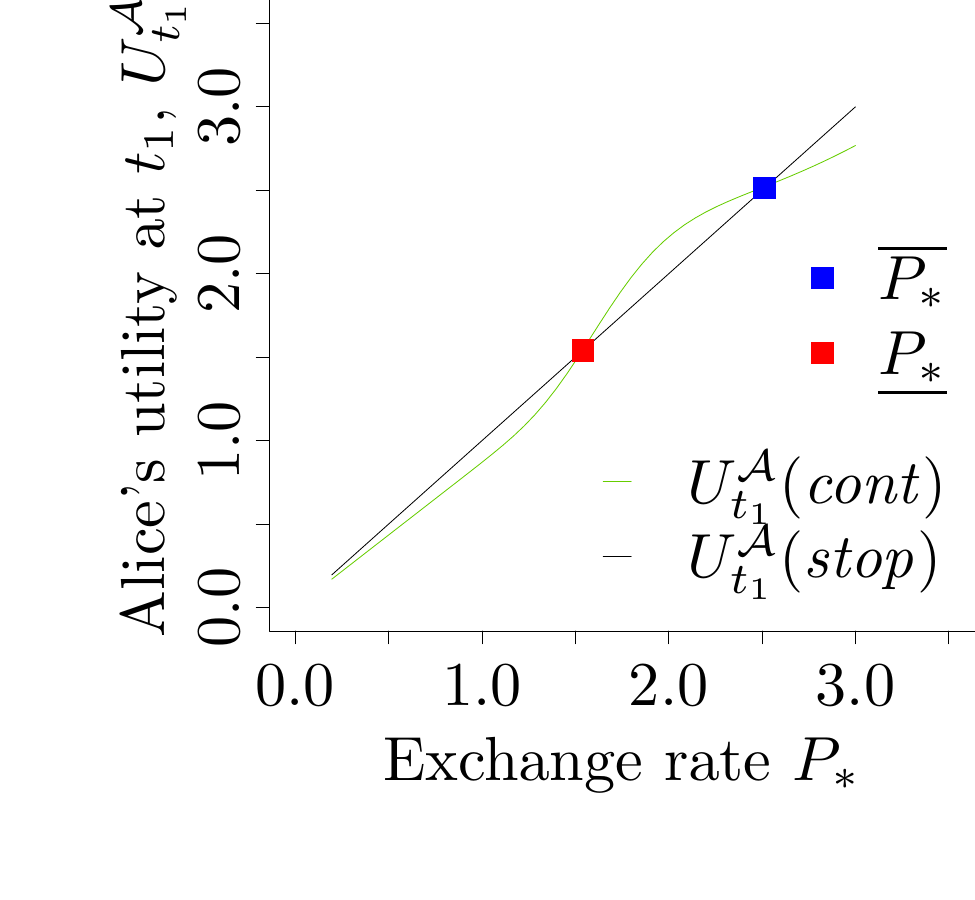}
    \caption{Alice's utility at $t_1$, ($U_{t_1}^\alice$) with different strategies ($\co, \st$), and exchange rate values ($P_*$).}
    \label{fig:util1_A}
\end{figure}

Thus, the exchange rate $P_*$ must lie within a range $(\underline{P_*}, \overline{P_*})$ to ensure the start of the swap (see \autoref{fig:util1_A}). Using the values from \autoref{tab:default} we numerically solve the feasible range as:\footnote{Note that $P_{t_1} = P_{t_0}$ since $t_1=t_0$ is assumed in \ref{eq:ideatime}.}
\begin{equation}
(\underline{P_*}, \overline{P_*}) = (1.5, 2.5)
\end{equation}

$\alice$'s strategy at $t_1$ can be summarized as:
\begin{equation}
    \begin{cases}
    \co, &  \underline{P_*} < P_* \leq  \overline{P_*}\\
    \st, & P_* \leq \underline{P_*} \text{ or } P_* > \overline{P_*}
    \end{cases}
\end{equation}

\subsection{Success rate}
\label{sec:sucrate}

We define the success rate ($\mathit{SR}$) of a swap to be the likelihood of completion of the swap {\em after} it has been initiated, i.e. after $\alice$ has made the first move at $t_1$. With the values of all other parameters being fixed (\autoref{tab:default}), $\mathit{SR}$ is a function of $P_*$, and can be expressed as:
\begin{align}
\mathit{SR}(P_*) = &
\int_{
\underline{P_{t_2}}(P_*)
}^{\overline{P_{t_2}}(P_*)
} \mathcal{P}\left(
x, P_{t_1},\tau_a
\right) \,
\Big[
1-
\mathcal{C}(\underline{P_{t_3}}(P_*) ,x,\tau_b)
\Big]
dx,
\nonumber \\
& \underline{P_*} < P_* \leq  \overline{P_*}
\end{align}

\begin{figure}
\centering
\includegraphics[
    height=0.0925\textheight,
    trim = {54, 105, 70, 5}, clip
    ]{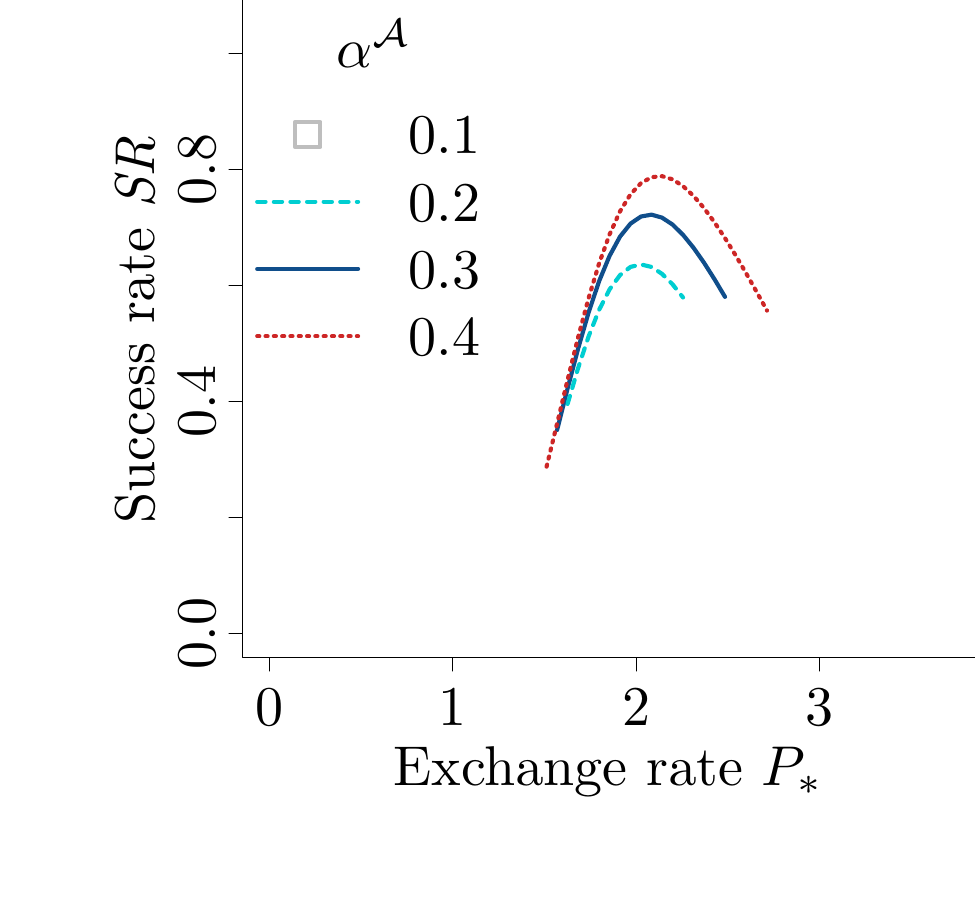}
~\includegraphics[
    height=0.0925\textheight,
    trim = {115, 105, 70, 5}, clip
    ]{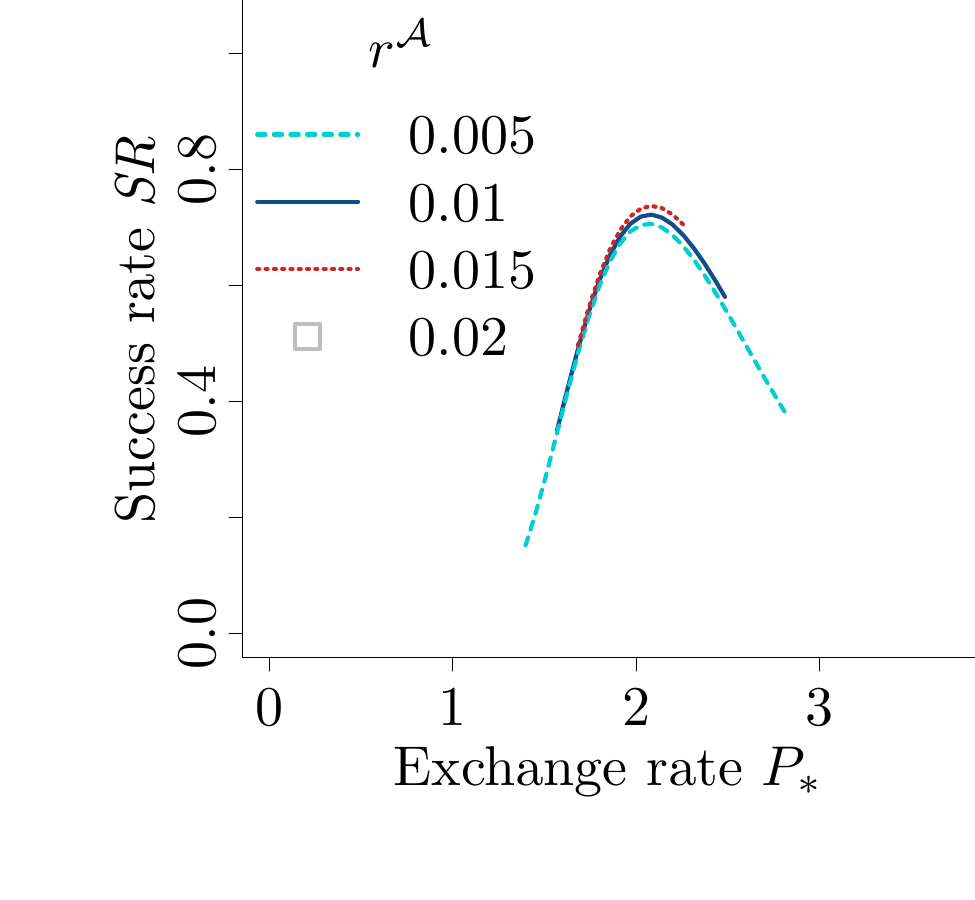}
~\includegraphics[
    height=0.0925\textheight,
    trim = {115, 105, 70, 5}, clip
    ]{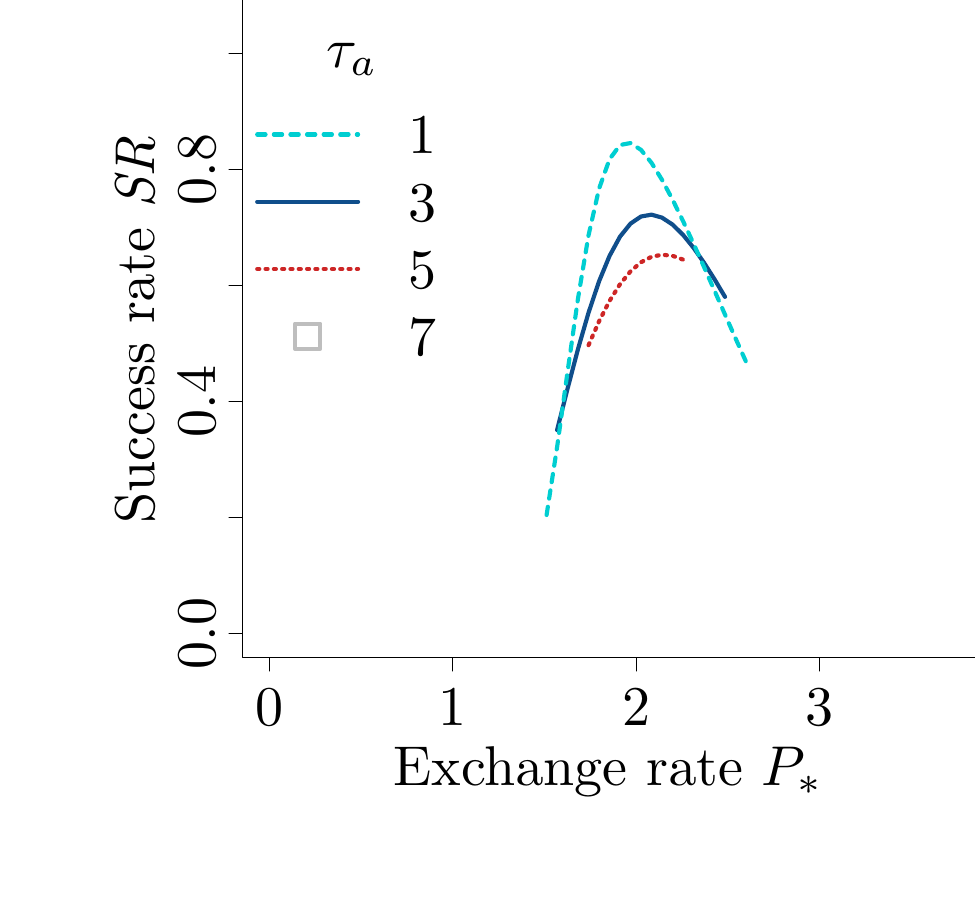}
~\includegraphics[
    height=0.0925\textheight,
    trim = {115, 105, 70, 5}, clip
    ]{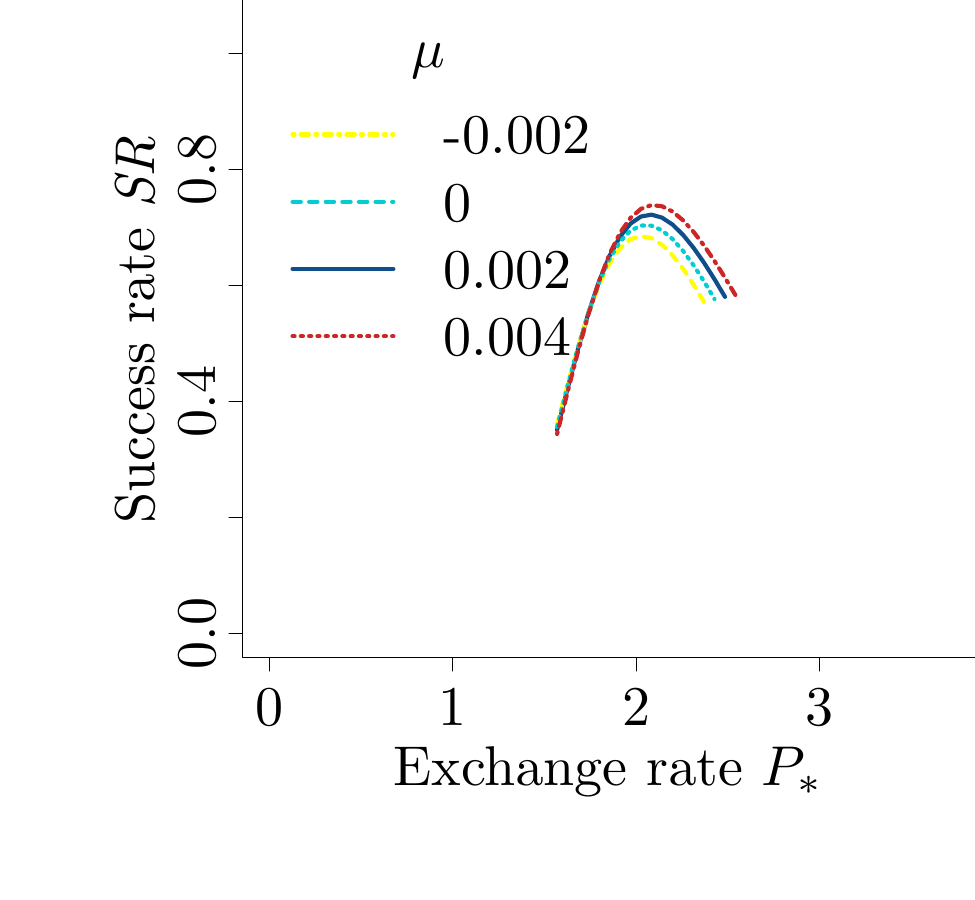}
    
\includegraphics[
    height=0.11\textheight,
    trim = {54, 47, 70, 5}, clip
    ]{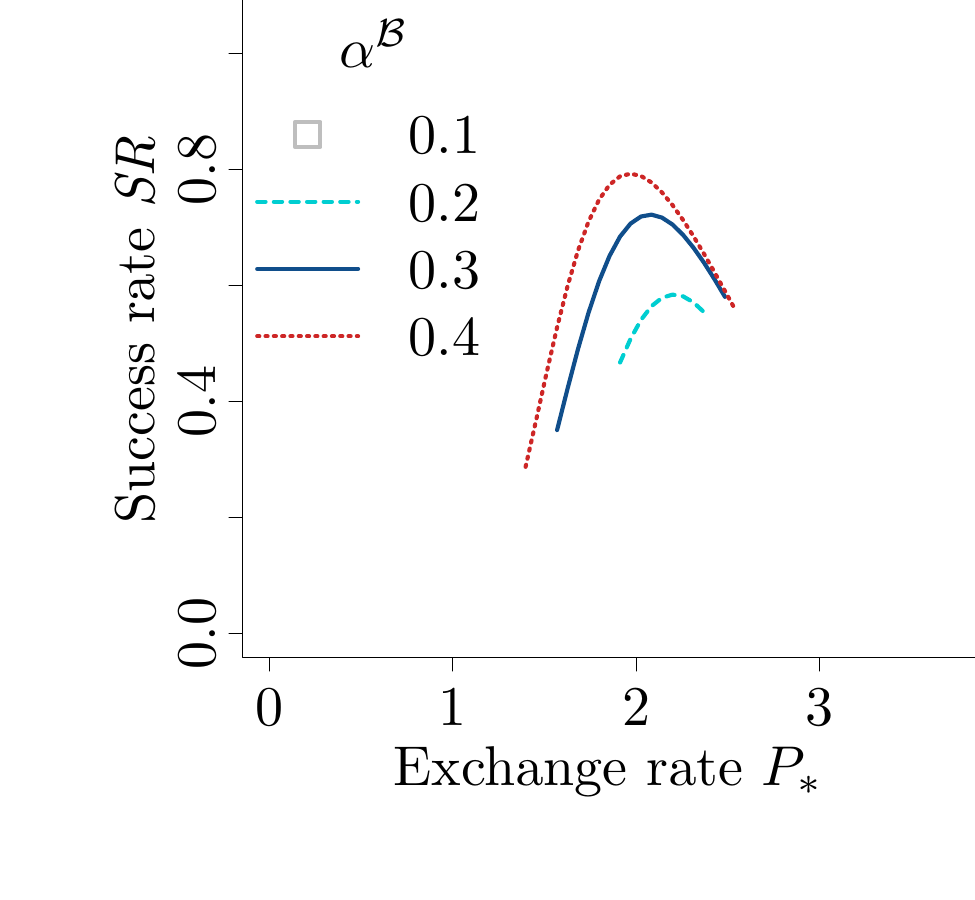}
~\includegraphics[
    height=0.11\textheight,
    trim = {115, 47, 70, 5}, clip
    ]{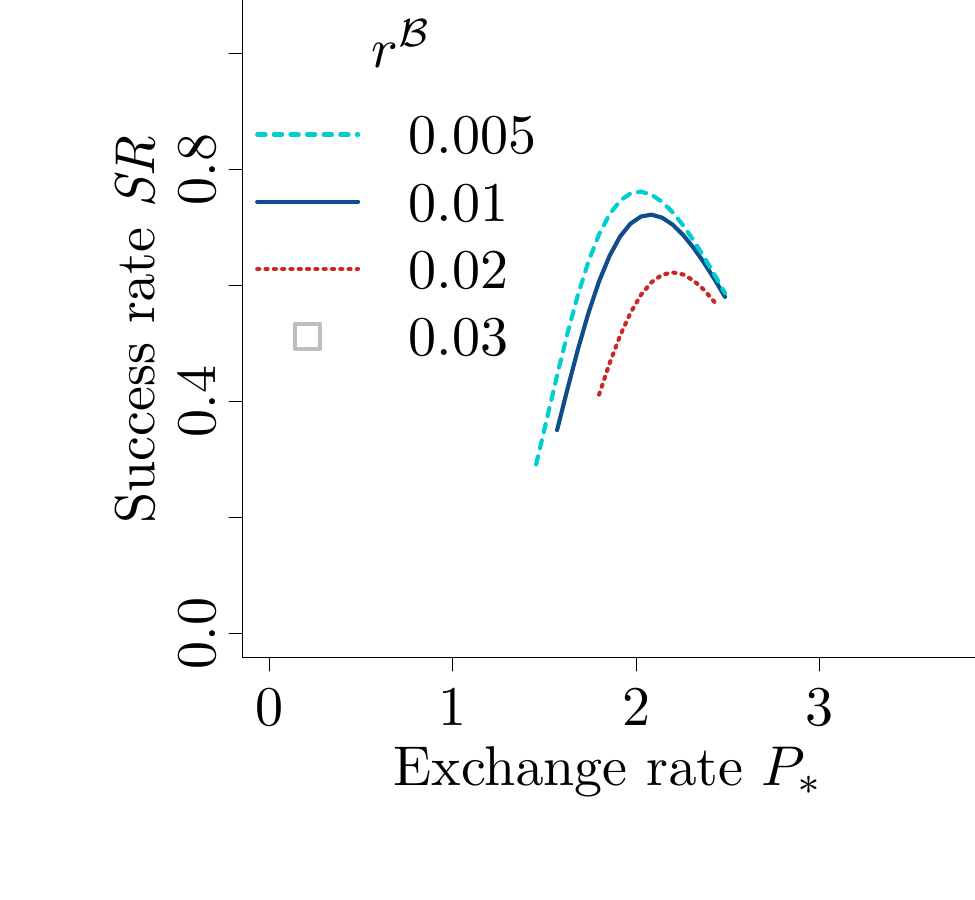}
~\includegraphics[
    height=0.11\textheight,
    trim = {115, 47, 70, 5}, clip
    ]{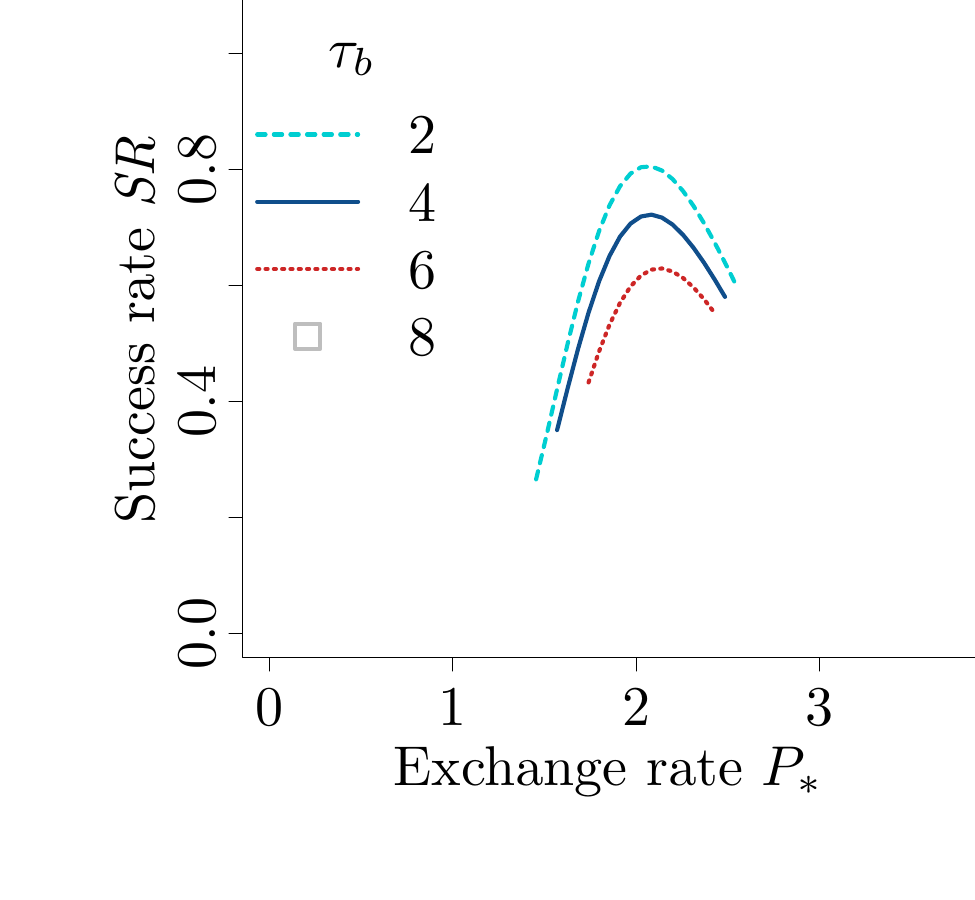}
~\includegraphics[
    height=0.11\textheight,
    trim = {115, 47, 70, 5}, clip
    ]{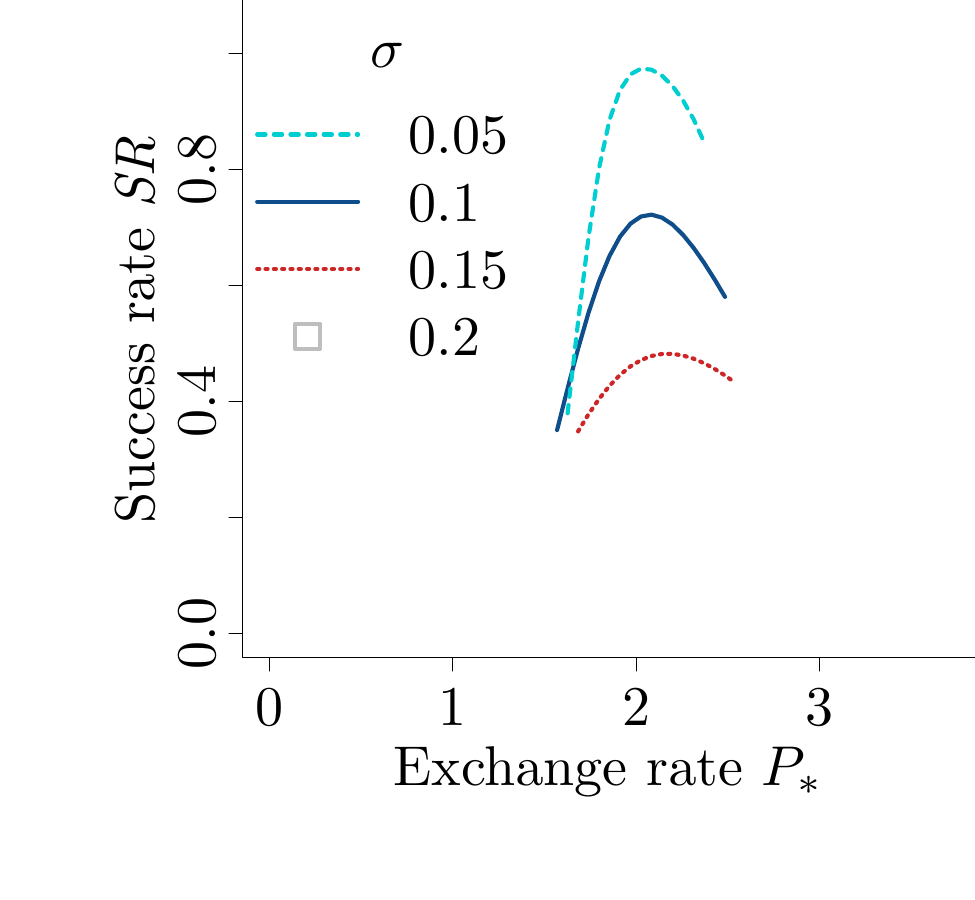}
\caption{Swap success rate $\mathit{SR}$ as a function of exchange rate $P_*$ with different parameter values. Default parameter values are set in \autoref{tab:default}.
Line plots illustrate $\mathit{SR}$ curves with viable parameter values; non-viable values are marked with
\textcolor{gray}{\tiny $\square$}.
}
\label{fig:sucrate}
\end{figure}

In \autoref{fig:sucrate}, we show how success rate $\mathit{SR}$ changes with the exchange rate $P_*$. $\mathit{SR}$ curves with the default parameter setting (\autoref{tab:default}) are plotted in blue line (\textcolor{dodgerblue4}{\myrule}), which are compared with $\mathit{SR}$ curves with different parameter values.
Irrespective of the parameter values, the $\mathit{SR}(P_*)$ curve is always concave, with the $\mathit{SR}$-maximizing point residing between $\underline{P_*}$ and $\overline{P_*}$. As suggested in Section \ref{sec:backwardation}, this is because overly low $P_*$ reduces the likelihood of continuation at $t_3$ and $t_2$, while overly high $P_*$ reduces the likelihood of continuation at $t_2$.

Next, we discuss how the value setting of other parameters affects the success rate.

\subsubsection{Success premium $\alpha$}
\label{sec:alpha}

The parameter success premium describes the excess utility that an agent receives when the swap succeeds.
The parameter captures not only the excess utility an agent gains from possessing the counterparty's token over his/her own token, but also the utility of guarding his/her reputation. That is to say, the more an agent cares about honoring an agreement, the higher $\alpha$ will be.
As shown in \autoref{fig:sucrate}, ceteris paribus, higher $\alpha$ leads to higher $\mathit{SR}$. This is true for both $\alpha^{\alice}$ and $\alpha^{\bob}$. In addition, higher $\alpha$ renders a bigger feasible range of $P_*$. Note that when $\alpha$ is too small (either with $\alice$ or $\bob$), the swap would never be initiated.

\subsubsection{Time preference $r$}

The parameter time preference describes an agent's {\em impatience} level. Larger $r$ suggests a higher degree of impatience, i.e. possessing an asset {\em right now} is more valuable for the agent than obtaining it {\em later}.
As an HTLC swap requires an asset-locking period, a certain degree of {\em patience} is needed for both $\alice$ and $\bob$ to enter the agreement. Thus, as shown in \autoref{fig:sucrate}, larger $r$ results in a narrower viable range of values for $P_*$; exceedingly high $r$ renders any $P_*$ value infeasible, i.e. the swap would never be initiated.

\subsubsection{Transaction confirmation time $\tau$}

With the presence of time preference ($r^\alice, r^\bob > 0$), longer transaction confirmation time, either on Chain$_a$ or Chain$_b$, reduces agents' utility in engaging in a swap. Therefore, higher $\tau_a$ or $\tau_b$ shrinks the viable range of $P_*$. When $P_*$ is always chosen optimally (as to maximize $\mathit{SR}$), lower $\tau_a$ or $\tau_b$ increases $\mathit{SR}$.

\subsubsection{Price trend $\mu$ and volatility $\sigma$}

\autoref{fig:sucrate} shows that, ceteris paribus, higher degree of upward price trend of Token$_b$ increases $\mathit{SR}$. In contrast, higher volatility reduces max $\mathit{SR}$.

\section{Model extension} \label{sec:extension}

In this section, we expand on our basic model described in Section \ref{sec:game} and discuss two derivations.

\subsection{HTLC with collateral}

\begin{figure}
$Q = 0.01$

\includegraphics[
    height=0.135\textheight,
    trim = {52, 40, 24, 0}, clip
    ]{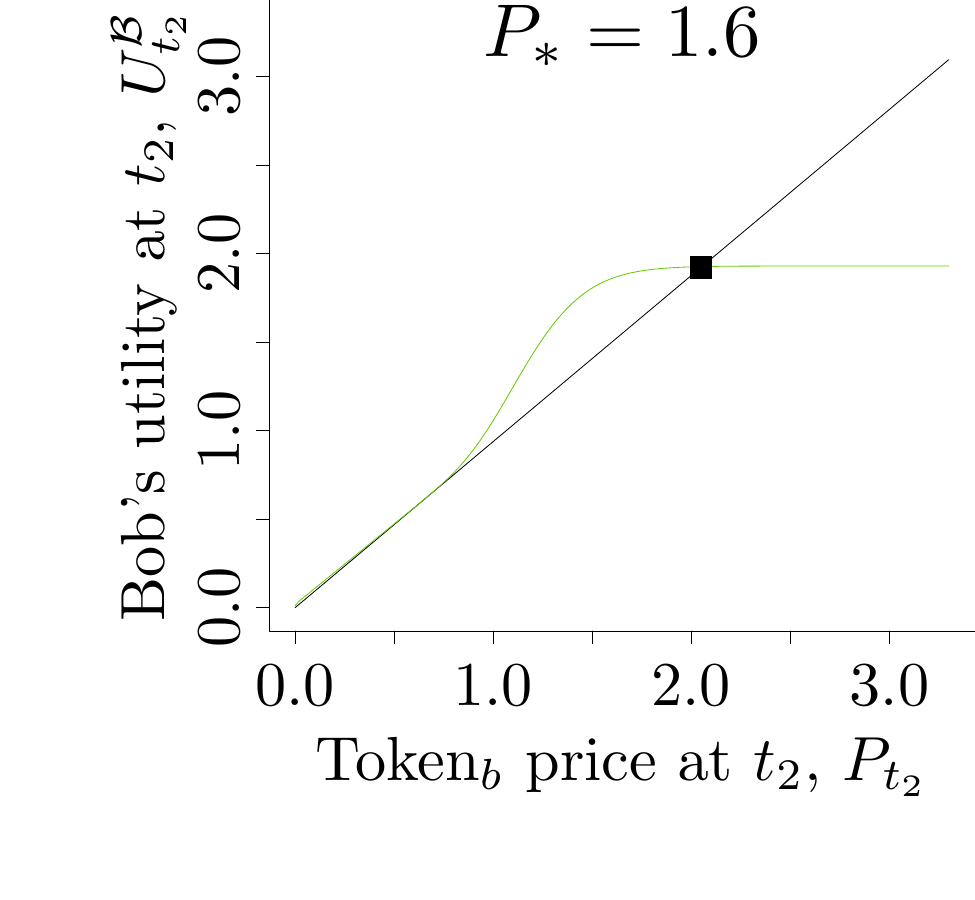}
~\includegraphics[
    height=0.135\textheight,
    trim = {120, 40, 24, 0}, clip
    ]{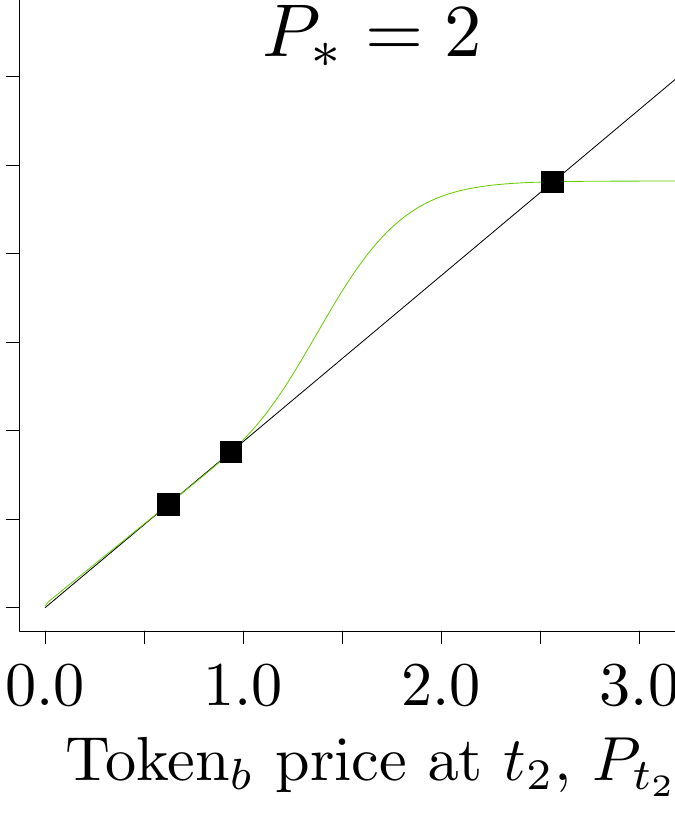}
~\includegraphics[
    height=0.135\textheight,
    trim = {120, 40, 15, 0}, clip
    ]{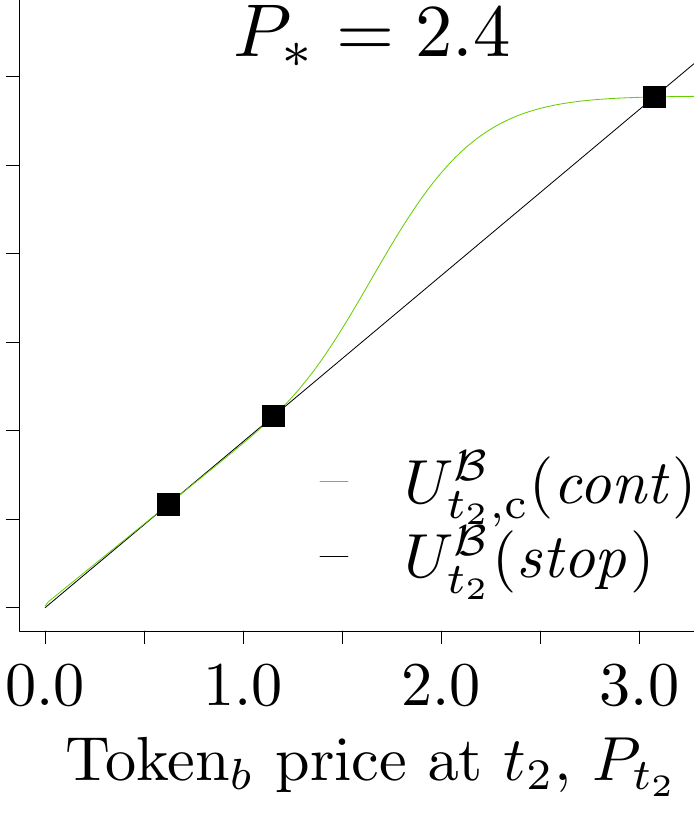}

$Q = 0.1$

\includegraphics[
    height=0.135\textheight,
    trim = {52, 40, 24, 0}, clip
    ]{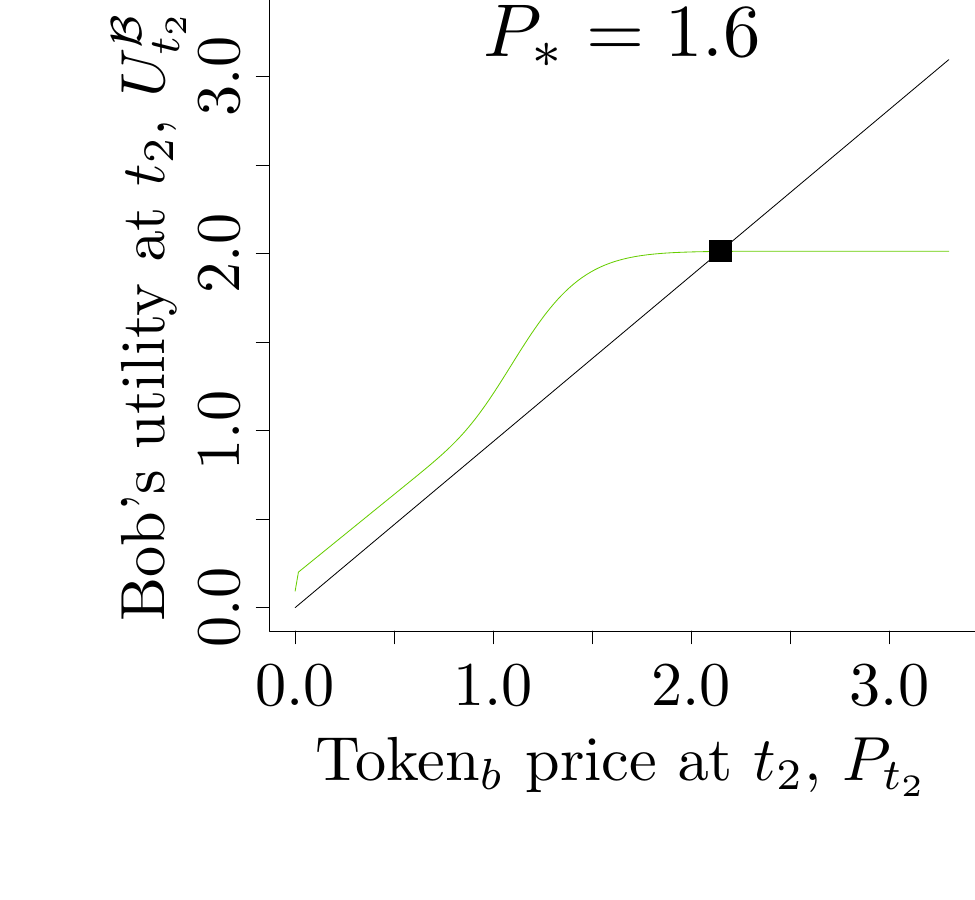}
~\includegraphics[
    height=0.135\textheight,
    trim = {120, 40, 24, 0}, clip
    ]{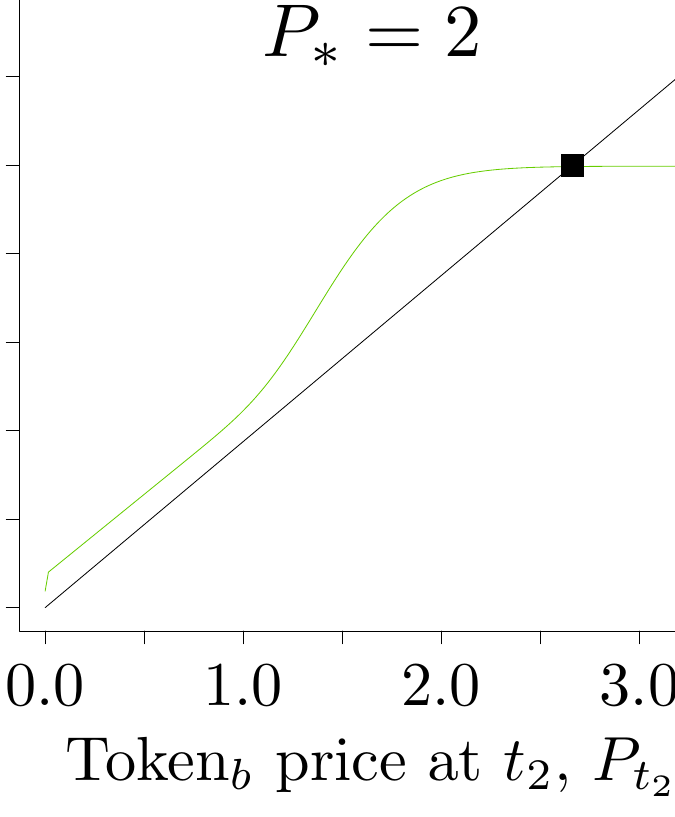}
~\includegraphics[
    height=0.135\textheight,
    trim = {120, 40, 15, 0}, clip
    ]{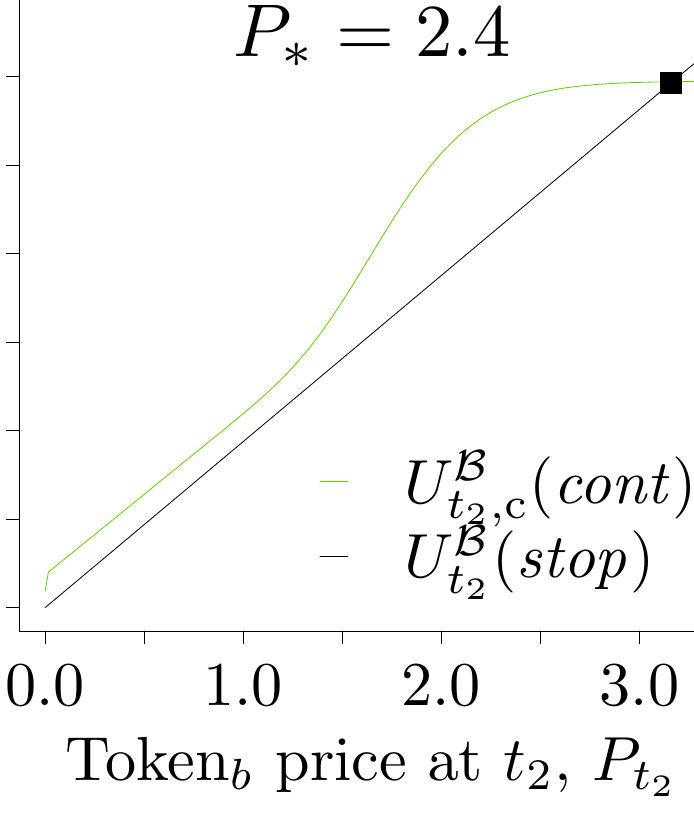}
    \caption{Bob's utility at $t_2$, ($U_{t_2}^\bob$). Indifference points between $\co$ and $\st$ are marked with {\tiny $\color{black} \blacksquare$}.}
    \label{fig:util2_B_col}
\end{figure}

In this section, we discuss an HTLC game where both agents place collateral into a smart contract before the actual swap.
All assumptions from Section \ref{sec:framework} with the exception of Assumption \ref{ass:utilityfunc} remain unchanged.
We assume additionally:

\begin{enumerate}
\item $\alice$ and $\bob$ move an allowance to a trusted smart contract on Chain$_a$ in order to charge each of them simultaneously the same amount of collateral, $Q$ Token$_a$, before the swap;
\item the smart contract is connected to an Oracle which observes the transaction outcomes on Chain$_a$ and Chain$_b$; \label{ass:oracle_obs}
\item if the swap succeeds, the Oracle transfers to each agent their original collateral; if an agent chooses $\st$ at any point during the swap, the other agent receives both agents' collateral from the Oracle; \label{ass:oracle_tra}
\item agent's utility function is as follows:
\begin{align}
U_{t,\colla}^i &= \mathbb{E}\left[
\tfrac{(1 + \alpha^i \, S) \,V_{t+T_t^i}}{
e^{r^i \, T_t^i}
} +
\tfrac{C_t^i}{
e^{r^i \, t_\colla}
}
\right]
\end{align}
where

$C$: value of collateral to be received back,

$t_\colla$: time until receiving the collateral.

\end{enumerate}

We use subscript ``$_\colla$'' only when an expression differs from the one in the basic setup (Section \ref{sec:backwardation}).

This setup is theoretical as there is presently no Oracle service that would be able to monitor the actions as described, to the best of our knowledge.
Yet, the Bisq framework is similar in spirit with the key difference that a human arbitrator replaces the Oracle.
The goal of this section is to study the impact of collateralization on the agents' behaviors and thus on the transaction outcome.
A new atomic swap protocol with collateral will be discussed in a follow-up work.

We again employ backward induction to derive agents' utility-maximizing strategy.

\subsubsection{$t_4$}
At this point, if $\alice$ has released the secret, the Oracle will determine that $\alice$ has fulfilled all her obligations and releases her collateral $Q$ Token$_a$ at $t_4$. Thus, $\alice$ will receive $Q$ Token$_a$ at $t_4+\tau_a$. Same as in the basic scenario described in Section \ref{sec:t4}, $\bob$ chooses to unlock Token$_a$ ($\co$) with certainty. If $\alice$ has not released the secret, the Oracle will transfer $\alice$'s collateral to $\bob$.

\subsubsection{$t_3$}
\label{sec:t3_col}
At this point, $\bob$ has written an HTLC on Chain$_b$ as agreed, and hence has no further chance for foul play. The Oracle thus return's $\bob$'s collateral, and $\bob$ receives $Q$ Token$_a$ at $\tau_3 + \tau_\alpha$.
If now $\alice$ waives the contract, then her utility equals $U^\alice_{t_3}(\st)$ as described in \autoref{eq:At3stop} and her collateral will be transferred to be in the next step. If $\alice$ chooses to unlocks the 1 Token$_b$, the swap succeeds and she receives Token$_b$ at $t_5$ plus her collateral $Q$ Token$_a$ at $\tau_4 + \tau_\alpha$.

Therefore, $\alice$ chooses $\co$ over $\st$ if:
\begin{align}
U^\alice_{t_3}(\st) & < U^\alice_{t_3}(\co) + \tfrac{Q}{e^{r^\alice (\varepsilon_b+\tau_a)}}
\nonumber \\
P_{t_3}  & >
\left(
\tfrac{P_*}{
e^{r^\alice(\tau_\epsilon + 2\tau_a)}
} -
\tfrac{Q}{e^{r^\alice (\varepsilon_b+\tau_a)}}
\right)
\tfrac{e^{(r^\alice - \mu) \tau_b}
}{1+\alpha^\alice}
\label{eq:p3l_col}
\end{align}

Since $P_{t_3} \geq 0$, we express $P_{t_3}$'s lower bound as:
$$
\underline{P_{t_3, \colla}}=
\tfrac{e^{(r^\alice - \mu) \tau_b}
}{1+\alpha^\alice}
\cdot \max\left(
\tfrac{P_*}{
e^{r^\alice(\tau_\epsilon + 2\tau_a})
} -
\tfrac{Q}{e^{r^\alice (\varepsilon_b+\tau_a)}}, \, 0
\right)
$$

\begin{figure}
\centering
$Q = 0.01$
\hspace{45pt}
$Q = 0.1
\vspace{-9pt}$

\includegraphics[
    height=0.135\textheight,
    trim = {50, 45, 10, 5}, clip
    ]{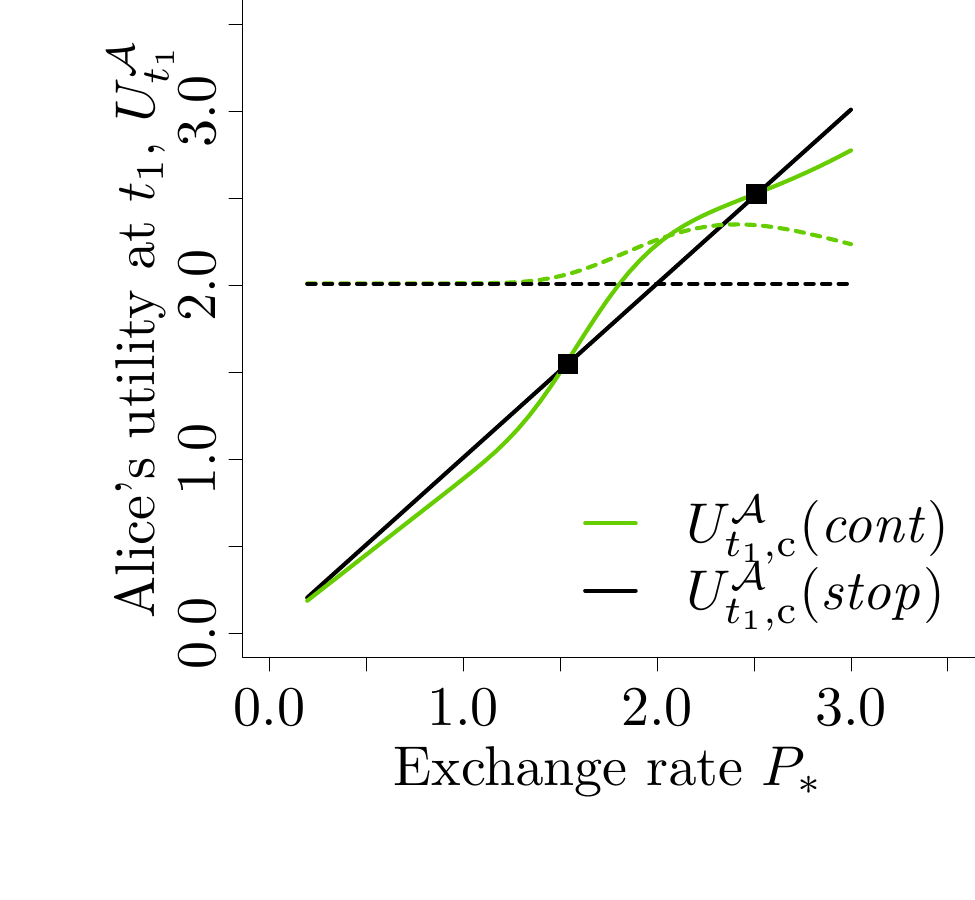}
~\includegraphics[
    height=0.135\textheight,
    trim = {110, 45, 10, 5}, clip
    ]{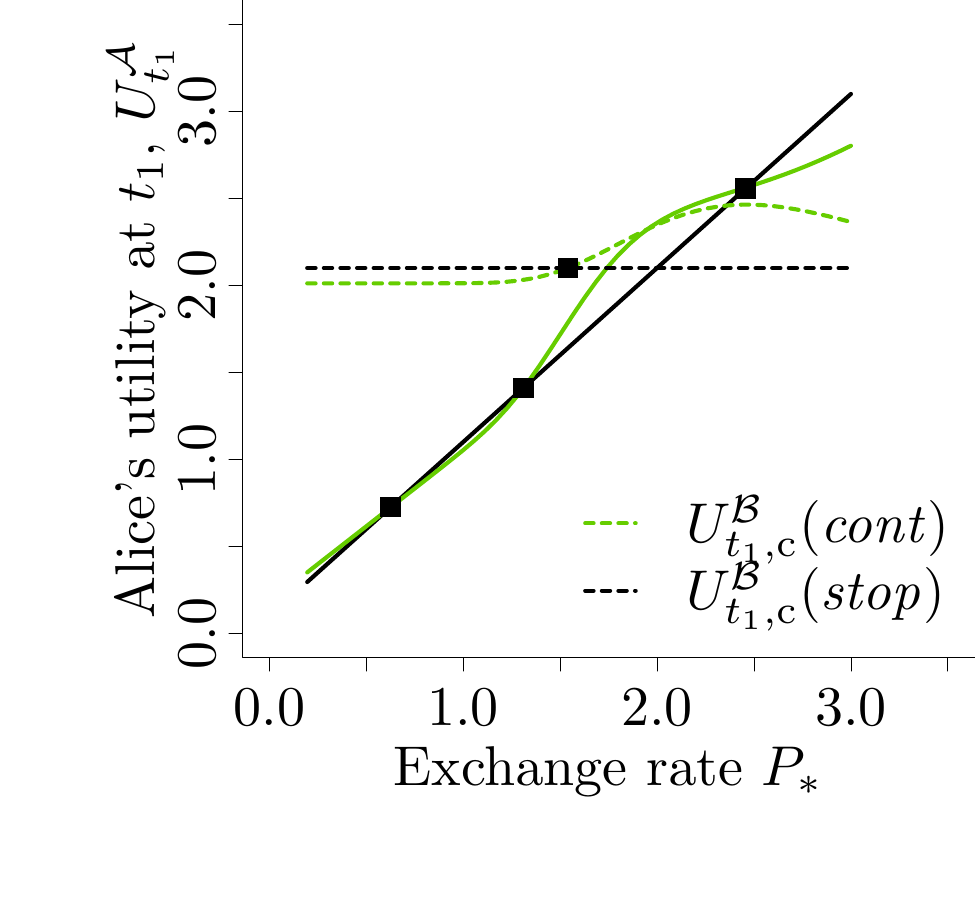}
\caption{Alice's and Bob's utility at $t_1$ ($U_{t_1}^\alice, U_{t_1}^\bob$). Each agent's indifference points between $\co$ and $\st$ are marked with {\tiny $\color{black} \blacksquare$}.}
\label{fig:t1AB}
\end{figure}

\subsubsection{$t_2$}
$\bob$ decides whether to write an HTLC on Chain$_b$ ($\co$) or not ($\st$).

\paragraph*{$\co$}
    As discussed in Section \ref{sec:t3_col}, if $\bob$ chooses $\co$ at this point, the Oracle will determine at $t_3$ that $\bob$ has fulfilled his obligations, and will return his collateral at that time.

    In addition, $\bob$ expects that at $t_4$, $\alice$ will honour the deal when $P_{t_3} > \underline{P_{t_3,\colla}}$ in which case $\bob$ gets $P_*$ Token$_a$ at $t_6$, and waive the deal otherwise in which case Bob gets 1 Token$_b$ at $t_7$ plus $\alice$'s collateral $Q$ at $t_3+\tau_a$.

Therefore, the utility of $\alice$ and $\bob$ utility at $t_2$ is:
\begin{align}
U^\alice_{t_2,\colla}(\co)
&
=
\tfrac{{
\int_{\underline{P_{t_3,\colla}}}^\infty
\mathcal{P}\left(x, P_{t_2},\tau_b\right)
\left(U^\alice_{t_3}(\co) + \frac{Q}{e^{r^\alice (\varepsilon_b+\tau_a)}}\right)
dx
\atop
+
\mathcal{C}\left(\underline{P_{t_3,\colla}}, P_{t_2},\tau_b\right)
U^\alice_{t_3}(\st)
}}{
e^{r^{\alice}\tau_b}
}
\\
U^\bob_{t_2,\colla}(\co)
& =
\tfrac{{
\frac{Q}{e^{r^\bob \tau_a}} +
\left[
1-\mathcal{C}(\underline{P_{t_3,\colla}}, P_{t_2},\tau_b)
\right]
U^\bob_{t_3}(\co)
\atop +
\int_0^{\underline{P_{t_3}}}
\mathcal{P}(x ,P_{t_2},\tau_b)
\left(
U^\bob_{t_3}(\st)
+ \frac{Q}{e^{r^\bob (\varepsilon_b + \tau_a)}} \right)
dx
}}{
e^{r^{\bob}\tau_b}
}
\end{align}

\paragraph*{$\st$}
$\bob$ withdraws from the deal and keep 1 Token$_b$.
$\bob$'s utility is the same as \autoref{eq:Bt2stop}.
The swap stops due to $\bob$'s foul play. The Oracle thus releases both agents collateral, $2Q$ in total, to $\alice$ at $t_3$, who will receive the fund at $\tau_3+\tau_a$.

Intuitively,
if $P_{t_2}$ is too high, then $\bob$ would like to keep the valuable Token$_b$ and wouldn't want to swap, and $\bob$ would therefore choose $\st$;
if $P_{t_2}$ is too low (say, close to zero), forfeiting the valuable collateral to keep the worthless Token$_B$ would not be sensible---even if $\alice$ chooses not to reveal secret in the next step, $\bob$ would at least be able to receive the collateral; therefore, $\bob$ would choose $\co$.

Thus, Equation $U^\bob_{t_2,\colla}(\co) = U^\bob_{t_2}(\st)$ has an \emph{odd} number of roots. \autoref{fig:util2_B_col} shows that there can be 1 or 3 intercepts between curve $U^\bob_{t_2,\colla}(\co)$ 
and $U^\bob_{t_2}(\st)$, depending on the value of $Q$ and $P_*$.
Define set $\mathfrak{P}_{t_2}$ such that:
$$
P_{t_2} \in \mathfrak{P}_{t_2} \iff
U^\bob_{t_2,\colla}(\co)>
U^\bob_{t_2}(\st)
$$
Hence, $\bob$ would choose $\co$ if and only if $P_{t_2}$ falls in $\mathfrak{P}_{t_2}$.

\begin{figure}
\centering
\includegraphics[
    height=0.135\textheight,
    trim = {50, 45, 20, 5}, clip
    ]{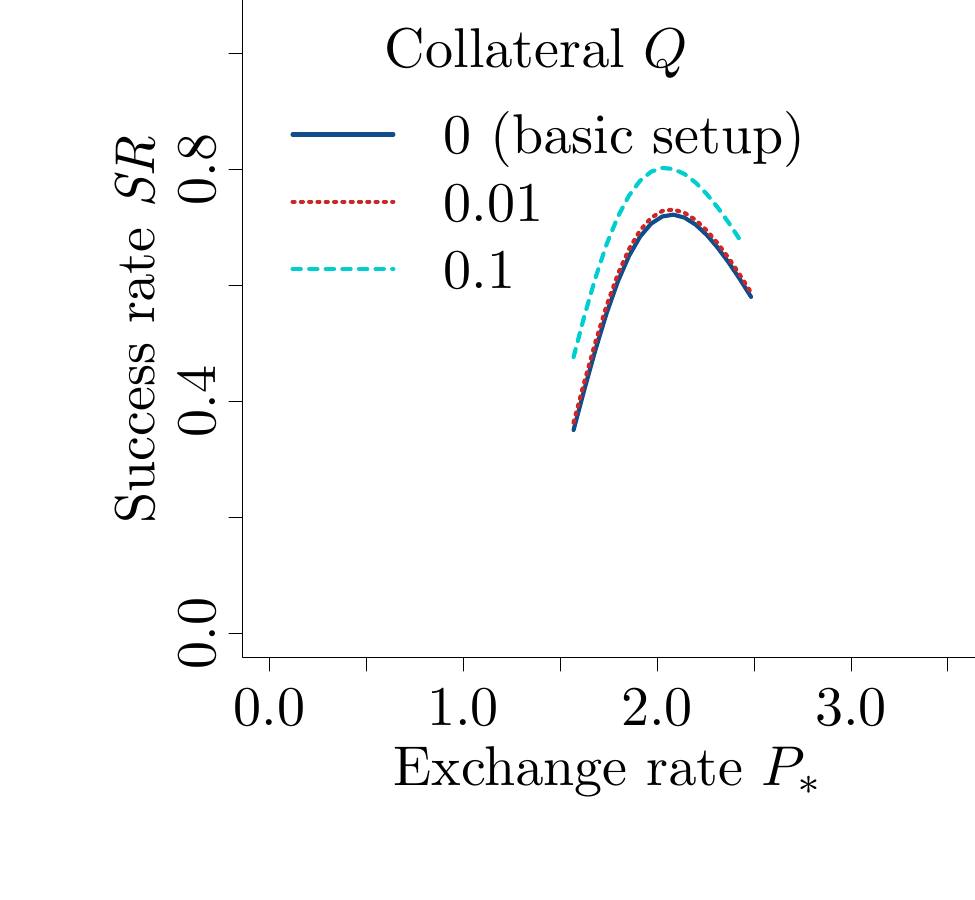}
\caption{Swap success rate $\mathit{SR}$ as a function of exchange rate $P_*$ with different collateral values $Q$.}
\label{fig:sucrate_col}
\end{figure}

\subsubsection{$t_1$}

$\alice$ and $\bob$ simultaneously make the decision on whether to engage in the swap ($\co$) or not ($\st$).

\paragraph*{$\co$}
The utility of $\alice$ and $\bob$ at $t_2$ can be expressed with their time-discounted, expected utility at $t_2 = t_1 + \tau_a$:

\begin{align}
U^\alice_{t_1,\colla}(\co)
&  =
\tfrac{{
\int_{
x \in \mathfrak{P}_{t_2}
}
\mathcal{P}(x ,P_{t_1},\tau_a)
U^\alice_{t_2,\colla}(\co)
dx +
\atop
\left(
\int_{
x \in \mathbb{R}^+ \setminus \mathfrak{P}_{t_2}
}
\mathcal{P}(x ,P_{t_1},\tau_a)
dx
\right)
\left(
U^\alice_{t_2}(\st) + \frac{2Q}{e^{r^\alice (\tau_b + \tau_a)}}
\right)
}}{
e^{r^{\alice}\tau_a}
}
\\
U^\bob_{t_1,\colla}(\co)
&  =
\tfrac{{
\int_{
x \in \mathfrak{P}_{t_2}
}
\mathcal{P}(x ,P_{t_1},\tau_a)
U^\bob_{t_2,\colla}(\co)
dx +
\atop
\int_{
x \in \mathbb{R}^+ \setminus \mathfrak{P}_{t_2}
}
\mathcal{P}(x ,P_{t_1},\tau_a)
U^\bob_{t_2}(\st) dx
}}{
e^{r^{\alice}\tau_a}
}
\end{align}

\paragraph*{$\st$}
$\alice$ and $\bob$ decide not to engage in the swap so that they can keep their original token and the collateral. Thus:
\begin{align}
U^\alice_{t_1, \colla}(\st) & =
U^\alice_{t_1}(\st) + Q = P_* + Q
\\
U^\bob_{t_1, \colla}(\st) & =
U^\bob_{t_1}(\st) + Q =
P_{t_1} + Q
\end{align}

Define set $\mathfrak{P}^\alice_{*}$ and $\mathfrak{P}^\bob_{*}$ such that:
\begin{align*}
P_{t_2} \in \mathfrak{P}^\alice_{t_2} \iff U^\alice_{t_2,\colla}(\co)>
U^\alice_{t_2,\colla}(\st) \\
P_{t_2} \in \mathfrak{P}^\bob_{t_2} \iff
U^\bob_{t_2,\colla}(\co)>
U^\bob_{t_2,\colla}(\st)
\end{align*}

Hence, the exchange rate $P_{*}$ must be in $\mathfrak{P}_{*} = \mathfrak{P}^\alice_{t_2} \cup \mathfrak{P}^\bob_{t_2}$, since otherwise agents' external utility ($\st$) exceeds the expected utility from the swap ($\co$) and the swap would not be initiated (see \autoref{fig:t1AB}).

The swap's success rate $SR$ can thus be expressed as:
\begin{flalign}
\mathit{SR}(P_*) =
&
\int_{
x \in \mathfrak{P}_{t_2}
} \mathcal{P}\left(
x, P_{t_1},\tau_a
\right) \,
\Big[
1-
\mathcal{C}(\underline{P_{t_3,\colla}}(P_*) ,x,\tau_b)
\Big]
dx,
\nonumber \\
&
P_* \in \mathfrak{P}_*
\end{flalign}

\autoref{fig:sucrate_col} shows $SR$ increases with collateral amount $Q$. This is because higher $Q$ allows for larger price movement, by expanding the feasible Token$_b$ price range at both $t_2$ (see \autoref{fig:util2_B_col}) and $t_1$ (see \autoref{eq:p3l_col}).

\subsection{Uncertain exchange rate}

\begin{figure}
\centering
\begin{subfigure}{0.49\linewidth}
\centering
\includegraphics[
	height=0.135\textheight,
	trim = {50, 45, 10, 5}, clip
	]{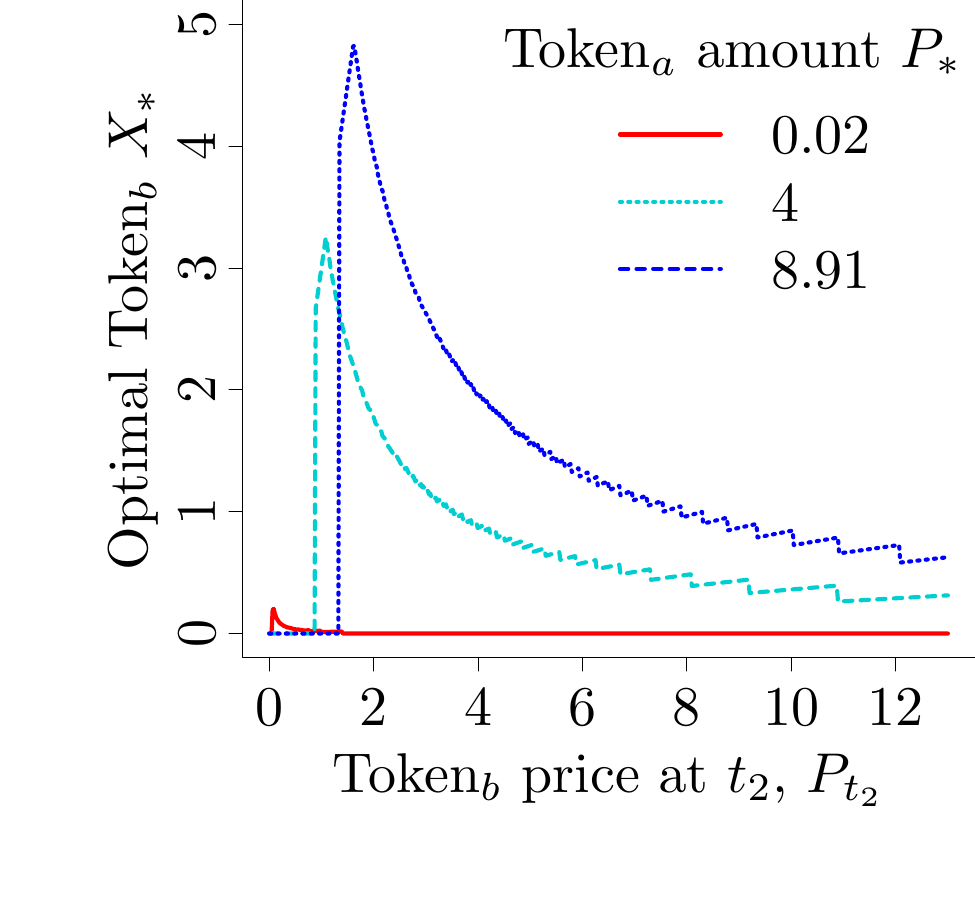}
\caption{Optimal Token$_b$ amount $X_*$ for Bob to lock based on $P_{t_2}$ and Token$_a$ amount $P_*$ locked at $t_1$.\label{fig:B_unknx}}
\end{subfigure}
\hfill
\begin{subfigure}{0.49\linewidth}
\centering
\includegraphics[
	height=0.135\textheight,
	trim = {50, 45, 10, 5}, clip
	]{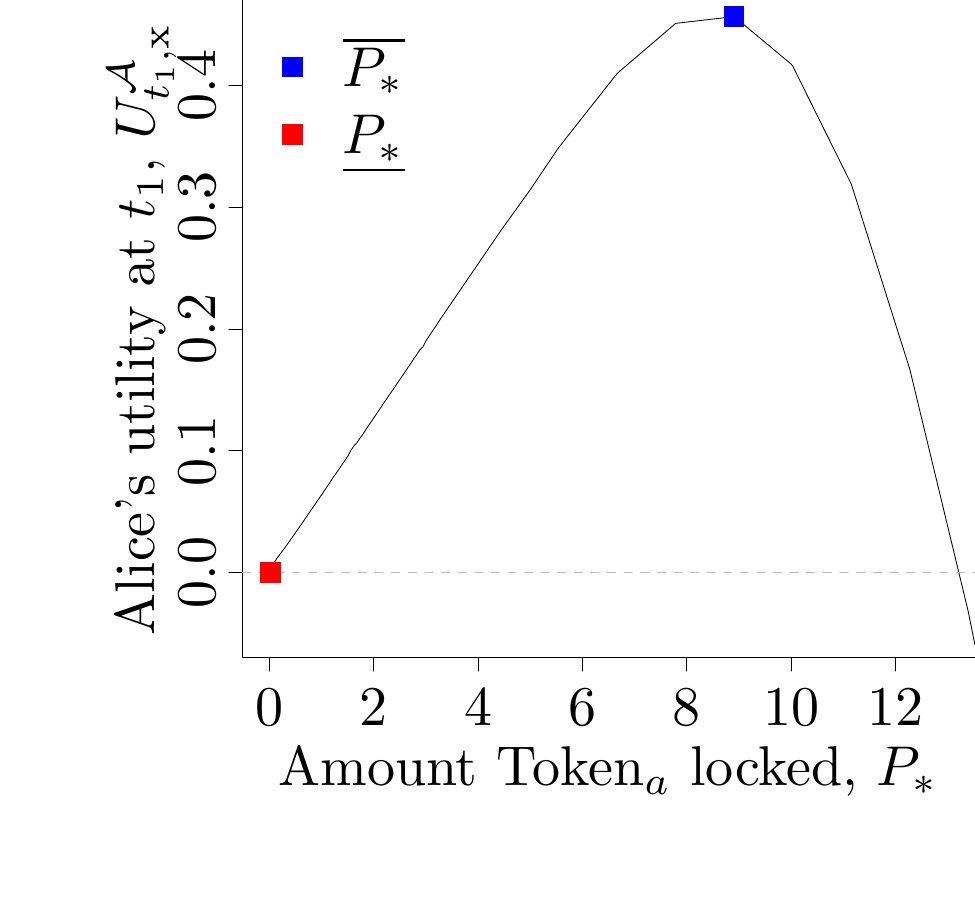}
\caption{Alice's utility at $t_1$, $U_{t_1,\unknx}^\alice$ as a function of the amount of Token$_a$ to lock $P_*$.\label{fig:A_unknx}}
\end{subfigure}
\caption{Agents' decision making with uncertain swap rate.}
\label{fig:AB_unknx}
\end{figure}

In this model variation, we discuss an HTLC game in which agents not only choose between $\co$ and $\st$, but also the exact amount of funds to lock in: $P_*$ Token$_a$ at $t_1$ and $X$ Token$_b$ $t_2$, respectively. This renders the actual exchange rate uncertain at the outset of the game.
We apply subscript ``$_\unknx$'' for expressions different from their counterpart in the baseline model.

\subsubsection{$t_4$}
Same as Section \ref{sec:t4}

\subsubsection{$t_3$}
Similar to Section \ref{sec:t3}, there exists a minimum feasible Token$_b$ price, expressed as $\underline{P_{t_3, \unknx}}(X)$ since it depends on $X$, the actual transferred amount at $t_2$. It can be easily derived that
\begin{align}
\underline{P_{t_3, \unknx}}(X) = \underline{P_{t_3}}/X,
\text{ with } \underline{P_{t_3, \unknx}}(0) = \infty
\end{align}
where $\underline{P_{t_3}}$ is expressed in Equation~\eqref{eq:p3lower}.

\subsubsection{$t_2$}
$\bob$ determines the value of $X \geq 0$ that maximizes his excess utility, namely the utility he obtains by proceeding with locking $X$ less the utility he keeps by retaining $X$ Token$_b$.
\begin{align}
U^\alice_{t_2, \unknx}(X)
& =
\tfrac{{
X
\int_{\underline{P_{t_3, \unknx}}(X)}^{\infty}
\mathcal{P}\left(x ,P_{t_2},\tau_b\right)
U^\alice_{t_3}(\co) dx
\atop +
\mathcal{C}(\underline{P_{t_3, \unknx}}(X) ,P_{t_2},\tau_b)
U^\alice_{t_3}(\st)
}}{
e^{r^{\alice}\tau_b}
}
\\
U^\bob_{t_2, \unknx}(X)
& = \tfrac{{
\left[
1-\mathcal{C}(\underline{P_{t_3, \unknx}}(X) ,P_{t_2},\tau_b)
\right]
U^\bob_{t_3}(\co)
\atop
+ X \int_0^{\underline{P_{t_3, \unknx}}(X)}
\mathcal{P}\left(x ,P_{t_2},\tau_b\right)
U^\bob_{t_3}(\st)dx
}}{
e^{r^{\bob}\tau_b}
}  - X P_{t_2}
\end{align}


Denote the optimal $X$ as $X_*(P_{t_2})$, a function of $P_{t_2}$. Thus,
\begin{align}
X_*(P_{t_2}) = \argmax_{X\geq 0} U^\bob_{t_2, \unknx}(X)
\end{align}

\autoref{fig:B_unknx} shows at any given amount of Token$_a$ that $\alice$ has locked at $t_1$, $P_*$, the optimal amount of Token$_b$ that $\bob$ should lock first increases and then decreases with $P_{t_2}$.

This aligns with the intuition that when $P_{t_2}$ is too low, i.e. Token$_b$ is valueless, then the likelihood of $\alice$'s withdrawal at $t_3$ is high, making $\bob$ unwilling to lock in big funds at $t_2$; when $P_{t_2}$ is too high, i.e. Token$_b$ is very valuable, then the amount of Token$_b$ that $\bob$ needs to commit also becomes low to make the deal worthwhile for himself. Generally at a given $P_{t_2}$, $X_*$ increases with $P_*$, reflecting a degree of fairness of the game.

\subsubsection{$t_1$}
$\alice$ takes into account the fact that $\bob$ will choose an amount of Token$_b$ to lock in at $t_2$ that maximizes his own utility, based on Token$_b$ price at $t_2$ and the amount of Token$_a$ that she commits, $P_*$.

$\alice$ thus chooses the value of $P_*$ that maximizes her excess utility, namely the utility she obtains by entering the swap with $P_*$ Token$_a$ in excess of the utility she keeps by retaining $P_*$ Token$_a$. Therefore,
\begin{align}
U^\alice_{t_1, \unknx}(P_{*})
=
\tfrac{
\int_{0}^{\infty}
\mathcal{P}(P_{t_2} ,P_{t_1},\tau_a)
U^\alice_{t_2, \unknx}(X_*(P_{t_2})) dP_{t_2}
}{
e^{r^{\alice}\tau_a}
} - P_*
\nonumber  \\
\label{eq:At1cont_unknx}
\end{align}

Similar to \autoref{fig:util1_A}, \autoref{fig:A_unknx} shows that the excess utility first increases and then decreases with $P_*$. $\overline{P_*}$ represents the amount that maximizes $U_{t_1,\unknx}^\alice$. Nevertheless, $\alice$ might only be able to afford a lesser amount due to a possible budget constraint on her side. $\underline{P_*}$ represents the lowest possible amount that $\alice$ needs to enter for a non-negative excess utility. 

\begin{figure}[t]
	\centering
	\includegraphics[
	height=0.135\textheight,
	trim = {50, 45, 20, 5}, clip
	]{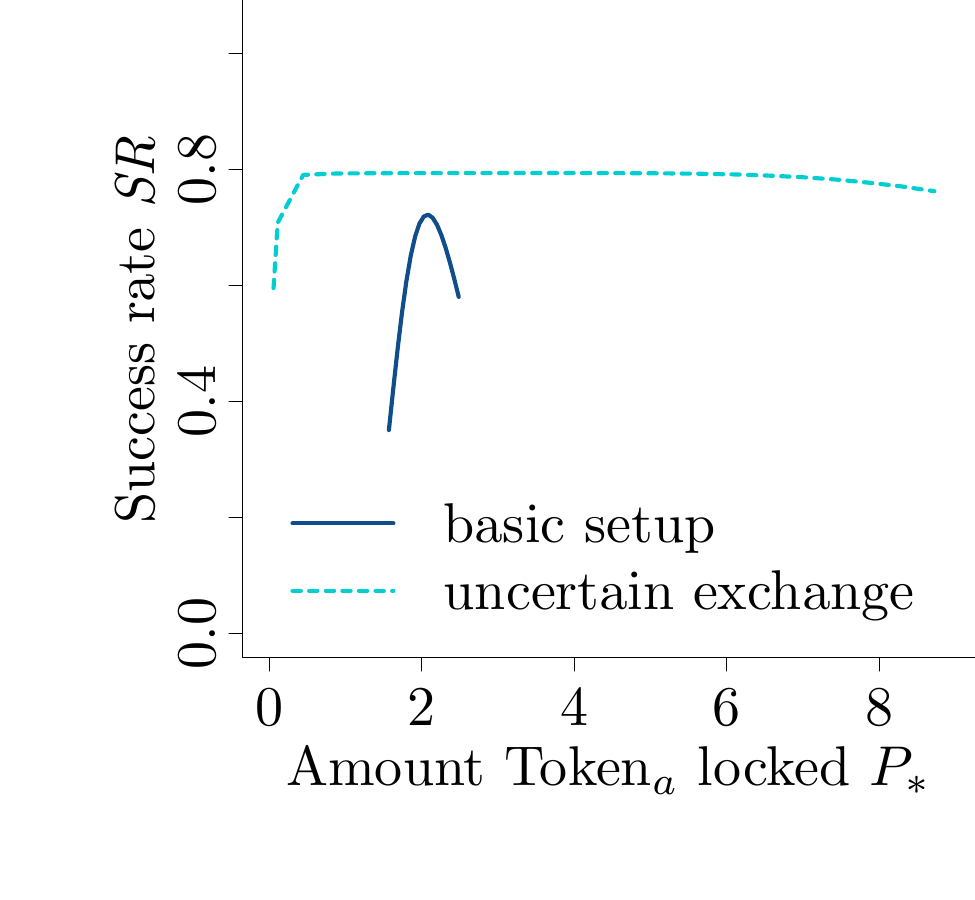}
	\caption{Swap success rate $\mathit{SR}$ as a function of initial amount of Token$_a$ locked in by Alice $P_*$.}
	\label{fig:sucrate_unknx}
\end{figure}

We can express the swap's success rate $SR$ under uncertain exchange rate as:
\begin{align}
	\mathit{SR}(P_*) = &
	\int_{
		0
	}^\infty \mathcal{P}\left(
	x, P_{t_1},\tau_a
	\right) \,
	\Big[
	1-
	\mathcal{C}(\underline{P_{t_3,\unknx}}(X(x)) ,x,\tau_b)
	\Big]
	dx,
\end{align}

\autoref{fig:sucrate_unknx} compares the success rate between the basic setup and the scenario with uncertain exchange rate. Interestingly, absence of pre-determined interest rate boosts the success rate.

\section{Discussion}

\subsection{Interpretation of findings}

Multiple findings can be drawn from our analysis that are relevant for real-world applications.
It has previously been mentioned that the agent completing the transaction receives a free \emph{American option}, meaning that she has the choice to complete the transaction, or not, based on whether the asset price changes at her advantage.
However, in this work, we show that the other agent (not only the swap initiator) may also leave the game midway, incentivized by a potentially higher financial gain. This scenario has thus far been neglected in the literature.



Our analysis also suggests that the collateral deposits can be dynamically adjusted depending on the terms of the swap (e.g. exchange rate) and optimization goal (e.g. maximizing utility, or maximizing success rate).

In the last model extension we show that the likelihood of completing the transaction is higher when the agents dynamically adjust the exchange rate to account for token price fluctuations. This is because an arbitrarily fixed exchange rate takes away the flexibility for agents to adjust their commitment based on the latest market condition (price change in our case).

\subsection{Limitations and future work}

Our work motivates multiple future research directions.

Firstly, simulation studies can be performed based on our model framework and its derivation using real market data.

Secondly, trustless protocols supporting collateral deposit without a third party can be designed.
To date, collaterals are typically deposited at a trusted third party in practice which is then responsible for resolving payment disputes. For instance, Bisq \cite{bisq}, an information platform for quotes and P2P transactions with arbitrators, uses a postage of collateral with possible intervention of an arbitrator, thus providing only a limited level of ``distributiveness'' and still requiring some trust in the arbitration system. 
Note that transactions executed via Bisq require a collateral deposit and an arbitrator fee.
Discussion with community members revealed that 3-5\% of transactions fail and go to arbitration, and that this percentage increases during periods of higher market volatility.

Thirdly, HTCL protocols can be further improved. 
HTLCs have known limitations \cite{Zamyatin, interoperabilityReview}, including strong assumptions required to maintain security, interactiveness, 
exclusiveness to public blockchains (for public mempools), and
the need for synchronizing clocks between blockchains and temporal locking of assets.

Lastly, a more realistic and sophisticated setup can be brought into our framework.
For example, future models may incorporate different risk-free rates for the two exchanged tokens, which resembles the settings of the Garman Kohlhagen model. In addition, blockchain transaction fees or coin stacking (similar to earning dividends or interest on a locked-in asset) may have an impact on agents' actions. Our model can also be extended to consider repeated games, stochastic individual utility, success premium as a random variable, etc.


\section{Conclusion}
\label{sec:conclusion}

We introduce a game-theoretic approach to model agent behaviors in cross-ledger transactions.
This allows us to study the viability and sensitivity of different protocols with respect to environment variables: agents' knowledge and utility functions, and the price dynamics.
We study in-depth the atomic swap as implemented with hash time lock contracts, for which we derived the success rate of a transaction.
In particular, we showed that both transacting counterparties can rationally decide to walk away from the transaction, and at different times.
This is a more realistic setup that relaxes the assumptions from previous works \cite{Han2019,diyhpl-eizinger} that only the swap initiator can benefit from price variation and act upon it.
A sensitivity analysis reveals that price volatility significantly affects the success rate of the transaction.

We extend our basic setup along two directions.
First, we show that introducing collateral deposit, in a purely theoretical way, increases the success rate of the transaction.
Second, we sketch out the scenario where agents are uncertain about the amount of funds their counterparty intends to commit.

Two important conclusions are therefore that cross-ledger atomic swap trustless protocols could benefit from the use of disciplinary mechanisms, such as collateral deposit, and that allowing agents to dynamically adjust the swap amount can increase the success rate.

\bibliographystyle{IEEEtran}
\bibliography{references}


\end{document}